\documentclass[5p, twocolumn, sort&compress]{elsarticle}
\makeatletter
\def\ps@pprintTitle{%
  \let\@oddhead\@empty
  \let\@evenhead\@empty
  \let\@oddfoot\@empty
  \let\@evenfoot\@oddfoot
}
\makeatother
\usepackage[english]{babel}
\usepackage[utf8]{inputenc}
\usepackage{geometry}
\usepackage{caption}
\usepackage{subcaption}
\usepackage{color, soul}
\usepackage{amsmath}
\usepackage{amssymb}
\usepackage{multirow}
\usepackage{array}
\usepackage{blindtext}
\usepackage{booktabs}
\usepackage{natbib}
\usepackage{graphicx}
\usepackage[colorlinks=true,allcolors=black]{hyperref}  

\usepackage[normalem]{ulem} 
\usepackage{xcolor}
\definecolor{orange}{rgb}{0.7,0.2,0}
\definecolor{darkgreen}{rgb}{0,0.5,0}

\newcolumntype{M}[1]{>{\centering\arraybackslash}m{#1}}

\journal{Carbon}

\begin{document}

\begin{frontmatter}

\title{Creation of nitrogen-vacancy centers in chemical vapor deposition diamond for sensing applications}

\author[1]{T. Luo}
\author[1]{L. Lindner}
\author[1]{J. Langer}
\author[1]{V. Cimalla}
\author[1]{F. Hahl}
\author[1]{C. Schreyvogel}
\author[2]{S. Onoda}
\author[2]{S. Ishii}
\author[2]{T. Ohshima}
\author[3,4]{D. Wang}
\author[3]{D. A. Simpson}
\author[5]{B. C. Johnson}
\author[6]{M. Capelli}
\author[7]{R. Blinder}
\author[1]{J. Jeske\corref{corr1}}
\ead{jan.jeske@iaf.fraunhofer.de}

\cortext[corr1]{Corresponding author. Tel: +49 761 5159-265}

\address[1]{Fraunhofer Institute for Applied Solid State Physics (IAF), Tullastr. 72, 79108 Freiburg, Germany}
\address[2]{National Institutes for Quantum Science and Technology (QST), 1233 Watanuki, Takasaki, Gunma 370-1292, Japan}
\address[3]{School of Physics, University of Melbourne, Melbourne, VIC 3010, Australia}
\address[4]{ARC Centre of Excellence in Exciton Science, School of Chemistry, University of Melbourne, Parkville VIC 3010, Australia}
\address[5]{School of Engineering, RMIT University, Melbourne 3000, Australia}
\address[6]{School of Science, RMIT University, Melbourne VIC 3001, Australia}
\address[7]{Institut für Quantenoptik, University of Ulm, D-89081 Ulm, Germany}

\begin{abstract}
The nitrogen-vacancy (NV) center in diamond is a promising quantum system for magnetometry applications exhibiting optical readout of minute energy shifts in its spin sub-levels. 
Key material requirements for NV ensembles are a high NV$^-$ concentration, a long spin coherence time and a stable charge state.
However, these are interdependent and can be difficult to optimize during diamond growth and subsequent NV creation. 
In this work, we systematically investigate the NV center formation and properties in chemical vapor deposition (CVD) diamond. 
The nitrogen flow during growth is varied by over 4 orders of magnitude, resulting in a broad range of single substitutional nitrogen concentrations of 0.2-20~parts per million.
For a fixed nitrogen concentration, we optimize electron-irradiation fluences with two different accelerated electron energies, and we study defect formation via optical characterizations. 
We discuss a general approach to determine the optimal irradiation conditions, for which an enhanced NV concentration and an optimum of NV charge states can both be satisfied.
We achieve spin-spin coherence times T$_2$ ranging from 45.5 to 549 $\mu$s for CVD diamonds containing 168 to 1~parts per billion NV$^-$ centers, respectively. 
This study shows a pathway to engineer properties of NV-doped CVD diamonds for improved sensitivity.
\end{abstract}

\begin{keyword}nitrogen vacancy center \sep%
    chemical vapor deposition \sep%
    eletron-beam irradiation \sep%
    magnetometry \sep%
    quantum sensing \sep%
    sensitivity
\end{keyword}

\end{frontmatter}

\section{Introduction}
The negatively charged nitrogen-vacancy (NV$^-$) center is a promising spin system for quantum sensing applications~\cite{aharonovich2011diamond,doherty2012theory,doherty2013nitrogen}.
Specifically, magnetic~\cite{maze2008nanoscale,mamin2013nanoscale,grinolds2013nanoscale,le2013optical} and electric~\cite{dolde2011electric,dolde2014nanoscale} field, strain~\cite{ovartchaiyapong2014dynamic,teissier2014strain}, temperature~\cite{acosta2010temperature,kucsko2013nanometre,neumann2013high,toyli2013fluorescence} and pressure~\cite{doherty2014electronic} sensing have all been demonstrated with the NV$^-$ center.

NV-ensembles with high NV$^-$ concentrations are favored for precision magnetometries~\cite{taylor2008high,acosta2009diamonds,barry2016optical,jeske2016laser}, for a considerably improved signal-to-noise ratio and sensitivity, which both ideally improve with the square root of the number of sensing spins~\cite{budker2007optical,barry2020sensitivity}.
High NV$^-$ concentrations require an enhanced nitrogen incorporation in diamonds, especially in the form of isolated-substitutional-nitrogen atoms (called P1 center or N$_s^0$ center)~\cite{lawson1998existence,jones2009acceptor}, which are converted to NV centers via irradiation and annealing.
However, the nitrogen impurities are also the main decoherence source of the NV spin, which limits the coherence time T$_2$~\cite{van1990optically,van1991fluorescence,hanson2008coherent} for quantum manipulation and sensing.
These competing parameters create a challenge when combining high NV$^-$ concentrations with long coherence times, which are both key factors for improved sensitivity in magnetic field sensing~\cite{rondin2014magnetometry,osterkamp2019engineering}. 
Consequently, the best achievable combination of NV$^-$ concentration and coherence time becomes an essential question for material optimization, which needs to be investigated throughout the diamond synthesis and after-growth treatments. 

Chemical vapor deposition (CVD) synthesis of diamond allows for well-controlled nitrogen-doping by setting the gas flow to the reaction chamber, typically at the lower end of nitrogen doping levels. High pressure high temperature (HPHT) synthesis, on the other hand, usually comes with larger nitrogen content (up to several hundreds of parts per million (ppm)) due to the contamination of the metal solvent-catalysts. 
However, HPHT diamonds often show strong variations in nitrogen densities between crystals grown under the same conditions, or even show sectors with vastly different P1 concentrations in a single crystal, due to the different nitrogen incorporation efficiencies along different growth directions during synthesis~\cite{capelli2019increased}.
Although both approaches have their advantages, the CVD synthesis is typically more controllable and reproducible: it can realize a high homogeneity of nitrogen incorporation due to growth in a single crystalline direction and a more precise control over the desired nitrogen density and thus NV coherence time.

In previous studies, limited ranges of nitrogen concentrations for specific interests have been mostly discussed, for example recently Edmonds~{\em et al.}~\cite{edmonds2021characterisation} have investigated CVD diamonds containing $\sim$10-15~ppm P1 centers and shown the reproducibility of their growth method by studying a large amount of samples; Rubinas~{\em et al.}~\cite{rubinas2021optimization} have studied HPHT diamonds with over $\sim$0.5-3~ppm P1 concentration range; and earlier Nöbauer~{\em et al.}~\cite{nobauer2013creation} have presented a study of HPHT diamonds with P1 densities of $\lesssim$200~ppm and CVD diamonds with P1 densities of $\lesssim$1~ppm.
In this work, we have grown CVD diamond series with a varying nitrogen flow over 4 orders of magnitude, leading to a broad range of P1 densities from 0.2 to 20~ppm.
To ensure comparability, we use a single reactor for all samples in this work.
We conducted systematic characterization on these CVD series to investigate P1, NV creations and NV charge states after growth, as well as their transformation under subsequent electron beam irradiation. 
Furthermore, we explore the optimal irradiation parameters, by growing two additional diamond series with a fixed initial P1 density of $\sim$2.2~ppm and irradiate them with multiple fluences at two different electron energies.

Conventional procedures for NV creation pursue mainly a high P1 to NV centers conversion and a high concentration of total NV centers.
The simultaneous reduction of the favourable NV$^-$ charge state often has been ignored.
Here through the irradiation, we study the increase in NV centers as well as the change in charge state distribution.
We aim to balance the increase of NV density with maintaining a high NV$^-$/NV ratio.
From this, we suggest general approaches to determine the optimal irradiation fluence for different nitrogen densities.

Moreover, we investigate the coherence time, T$_2$, of the NV center after growth, as well as after irradiation and annealing. 
We show the connection between P1 center density and T$_2$ after growth and achieve NV ensembles with long T$_2$. 
Through irradiation and annealing treatments, we monitor the change in T$_2$ and aim to increase the NV density without compromising T$_2$. 
We discuss combinations from high NV$^-$ with short T$_2$, to low NV$^-$ with long T$_2$, which help to identify the appropriate combination to achieve the optimal sensitivity for different applications.

\section{Materials and methods}

\subsection{Sample processing}
\label{chap:samples}
\begin{figure*}[!ht]
    \centering
    \includegraphics[width=0.85\textwidth]{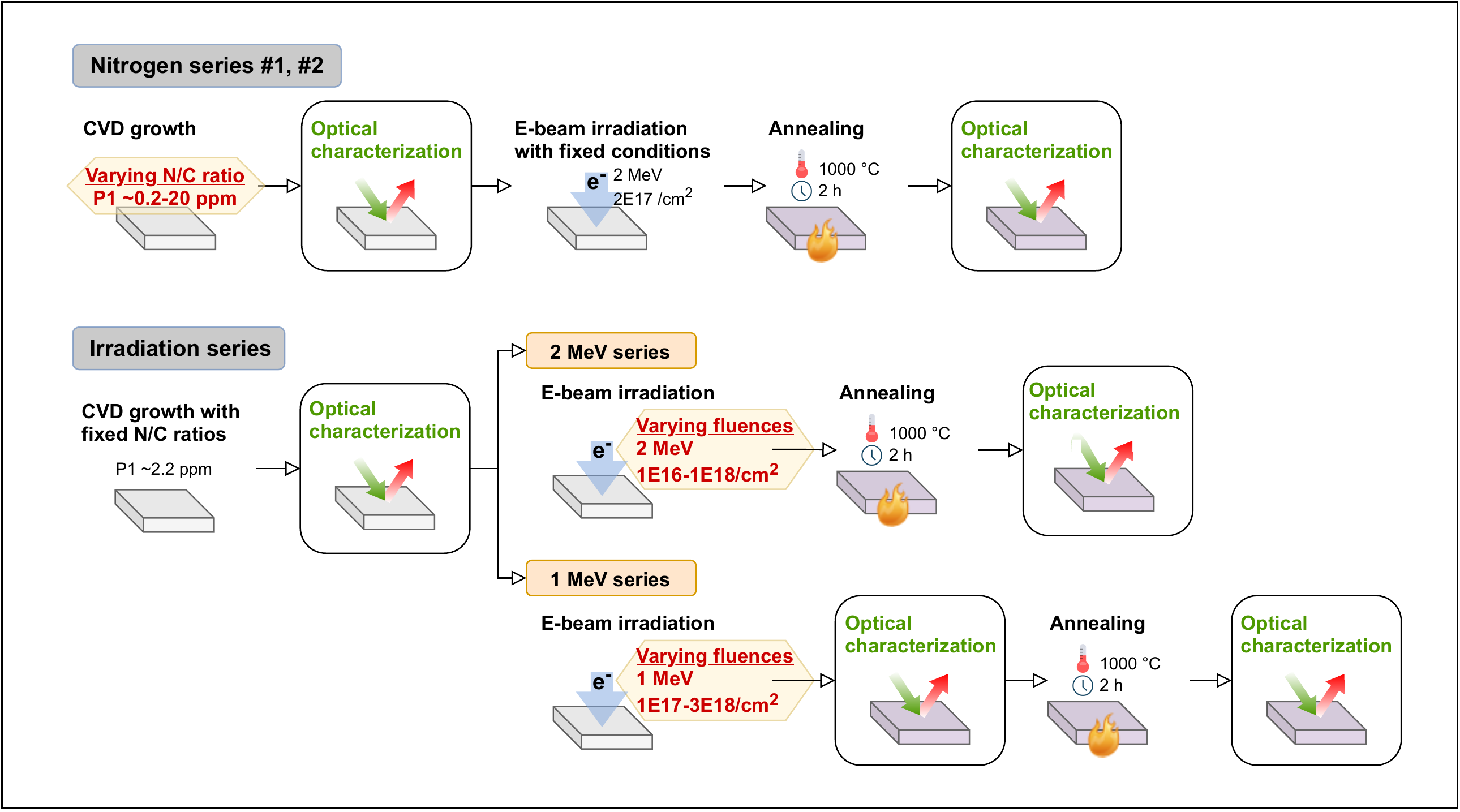}
    \put(-435,227){a)}
    \put(-435,142){b)}
    \caption{Flowchart of the sample preparation and characterization. a) The `nitrogen series' were grown with varying nitrogen concentrations, then irradiated and annealed under the same conditions. b) The `irradiation series' were grown with fixed nitrogen concentration, then irradiated under different conditions. Characterizations have been conducted after growth, then after irradiation and annealing steps.}
    \label{fig:floatchart}
\end{figure*}

In this study, we aim to systematically investigate the growth and irradiation steps, in order to study the NV center and P1 center creation and corresponding NV-T$_2$ of CVD diamond.
To this end, we have grown two sets of (100) orientated CVD diamonds.
These are referred to as the `nitrogen series', with nitrogen flow as the parameter and fixed irradiation conditions, and the `irradiation series' with fixed nitrogen flow and irradiation fluence as the parameter, Fig.~\ref{fig:floatchart}. 
The `nitrogen series' were grown to study the in-situ creation of P1 and NV centers, the NV charge state distribution, and the correlation between the P1 concentration and the coherence time.
Nitrogen series~$\#$1 was grown with lower concentrations of nitrogen added to the precursor gas, and Nitrogen series~$\#$2 with higher ones.
To more precisely tune the high concentration of nitrogen addition, a cascade of mass flow controllers was installed to the CVD reactor to study higher nitrogen flows.
The `irradiation series', then were studied to optimize the e-beam irradiation, therefore we have grown two irradiation series, which were irradiated with different electron energies and are respectively named the 2~MeV- and 1~MeV-series. 
All samples in both irradiation series were grown under the same condition to obtain samples with the same initial P1 concentrations. 
They were then irradiated with different fluences. 
Based on the study of the irradiation series, we irradiated the Nitrogen series~$\#$1 with the fixed optimal fluence to investigate how different nitrogen levels behavior under the same irradiation fluence.
The annealing treatment after irradiation was conducted at 1000~$^{\circ}$C for 2h for all samples.
Optical characterizations have been conducted after growth and after irradiation and annealing steps. 

The nitrogen-doped CVD growth of all samples was run in an ellipsoidal-shaped CVD reactor with a 2.45~GHz microwave frequency and equipped with a 6~kW microwave generator~\cite{funer1998novel}. 
The growth series were carried out at 210~mbar, between 800-900~$^{\circ}$C on the substrate measured by a narrow band radiation thermometer, with 2.2–2.7$\%$ of methane, with 0.1~sccm oxygen and an adjustable flow of nitrogen doping gas.

\subsection{Characterization methods}
\label{chap:method}
The P1 concentration for the nitrogen series was assessed with X-band continuous wave electron paramagnetic resonance (EPR) spectra collected at room temperature.
The P1 concentration for Nitrogen series~$\#$1 and ~$\#$2 were measured respectively by the spectrometer Bruker ELEXSYS E500 and Bruker ELEXSYS E580, both fitted with a Bruker super-high-Q probehead (ER4122 SHQE).
The microwave frequencies were set to 9.45~GHz and 9.84~GHz respectively.
For the Nitrogen series~$\#$1, before each measurement, a HPHT diamond (secondary calibration) sample with a known spin number was measured to quantify the spin number in each sample.
For the Nitrogen series~$\#$2, determination of the spin concentrations was performed using the built-in spin-counting feature, from the acquisition software (xEPR).
For some sample orientations, it was found that the cavity could not be properly critically coupled. We attributed this effect to the fact that sample edges, made of graphite-like carbon, are conducting. Proper sample orientation was chosen to enable critical coupling. This consists in putting two opposite edges of the square samples parallel to the sample insertion axis. For that cavity, such orientation ensures that no current loop will be created, because of the perpendicularity of the electric microwave field (TE102 mode) to two of the sample edges.
The EPR measurement for P1 concentration carries a $\sim$6$\%$ error.

The P1 concentration of the irradiation series was estimated from the absorption spectrum (measured with PerkinElmer Lambda 950) in the UV-Visible range (200-800~nm), in which an absorption band at 270~nm has been assigned to the P1 center~\cite{khan2009charge,dobrinets2016hpht}. To obtain the P1 concentration from this band, we compared its strength to reference samples with known P1 concentration measured by EPR. The main error was induced by calibration from EPR method.

 \begin{figure}[h!]
    \centering
    \includegraphics[width=\linewidth]{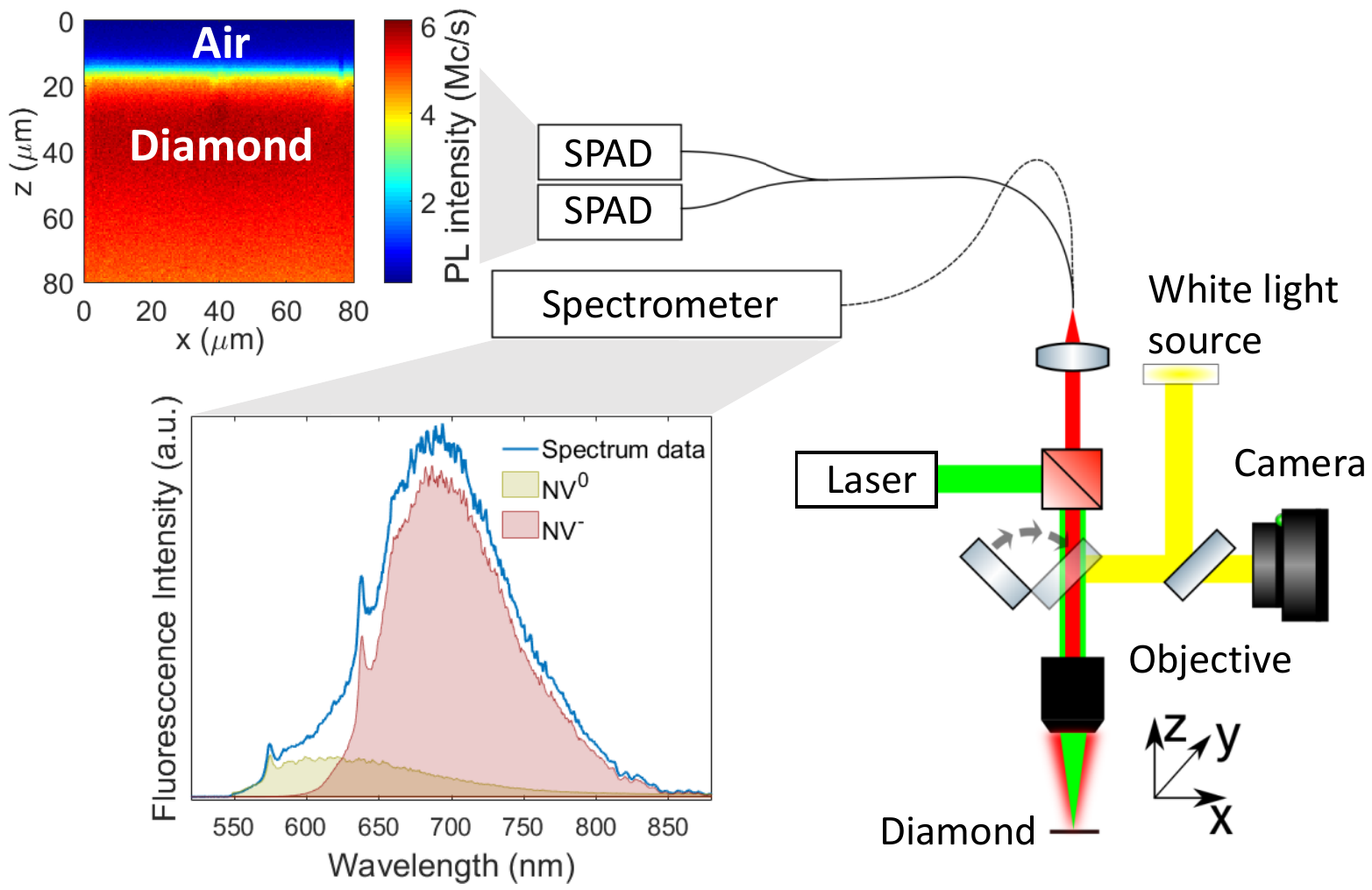}
    \caption{Schematic for the Photoluminescence (PL) measurement of NV-doped diamond plates}
    \label{fig:Sp_separation}
\end{figure}

The NV concentration was estimated from photoluminescence (PL) maps and spectroscopy by a home-built confocal setup (Fig.~\ref{fig:Sp_separation}): NV-diamonds were excited with a 532~nm laser (CrystaLaser), with a power of 10~$\mu$W, their PL was detected by single-photon avalanche diodes (Laser Components Count-T series) and a spectrometer (HORIBA iHR320).
The NV concentration ($[NV]$) from the PL signal was calibrated by the UV-visible absorption spectrum with the relation~\cite{capelli2019increased}: $[NV]=\mu_{532}/\sigma_{532}$, where $\sigma_{532}=(0.95\pm0.25)\cdot10^{-16}$~cm$^2$ is the absorption cross-section at 532~nm for NV centers~\cite{chapman2011quantitative}, and $\mu_{532}$ is the absorption coefficient (based on the exponential rate, not the decadic rate) of the diamond at the same wavelength. 
The PL measurement has high accuracy to compare NV emission between different samples, but the conversion to absolute concentration values carries $\sim$30\% error, which is induced by the absorption cross-section's literature value error.
The concentration of each NV charge state is determined from the NV emission spectrum, by fitting a weighted sum of the reference spectra of both NV charge states to it, determining the weighting factors that fit the original spectrum best by a least-squares fit. 
The NV$^0$ reference was taken with a shorter excitation wavelength (450~nm), the NV$^-$ reference was taken from a sample with only NV$^-$ (at the excitation power of 10~$\mu$W). 
To then determine the NV$^-$/NV$^0$ ratio, the photon number ratios ($N_p^-/N_P^0$) were corrected with the decays rates $\Gamma^-=1/12$~ns$^{-1}$ and $\Gamma^0=1/20$~ns$^{-1}$ respectively for the two charge states~\cite{liaugaudas2012luminescence,zhang2017dependence}, with the relation: $[NV^-]/[NV^0]= (N_p^-/N_P^0)\cdot(\Gamma^-/\Gamma^0)$.
Absolute NV$^-$ and NV$^0$ concentrations were then calculated as fractions of the total NV concentrations.

The coherence time, T$_2$, was measured by a home-built widefield microscope.
A modulated 532~nm excitation laser (Laser Quantum opus) was focused onto the samples at a power of 100~mW with a spot size of 0.25~mm, then the NV-PL from a small central region of (16$\times$16$~\mu$m) was collected by a sCMOS camera (ANDOR ZYLA 5.5 sCMOS). 
The T$_2$ measurement was conducted using the Hahn-echo protocol with MW $\pi$ time of 120~ns and readout laser time of 8.5~$\mu$s and repeated for 20~ms. The initial laser polarization time was 50~ms and background magnetic field was 700~G. The microwave was delivered via an omega-shaped gold resonator with inner diameter of 0.3~cm. The collected PL count was referenced against an immediate secondary spin-echo measurement with an addition $\pi$-pulse and T$_2$ is extrapolated from the exponential function $f(t)=a\cdot exp(-t/T_2)+c$.

\section{Results and discussion}
\subsection{As-grown NV-doped CVD diamond}
Nitrogen-doped CVD diamonds usually show detectable NV-fluorescence after growth, since some NV centers are formed directly in the growth, and the growth temperature activates local vacancies to combine with P1 centers but does not reach the point that NV centers become unstable. 
Working with NV centers that are formed directly during CVD growth is appealing for its technological simplicity as it does not require any irradiation or annealing steps. 
In this section we first study the in-situ creation of NV centers as well as the incorporation of P1 centers. 
We have grown the two nitrogen series to map concentrations of P1 centers and NV centers created by different nitrogen flows, showing how NV centers correlate to P1 centers; and how NV charge states behave in this phase. 
Furthermore, we investigate the coherence time of the series, and plot it as a function of EPR-determined P1 center densities. 
With all these characterizations we discuss the performance of nitrogen-doped CVD diamonds as a sensing material, without further treatments.

\subsubsection{P1 and NV creation as grown}
\label{chap:NV_asgrown}
\begin{figure*}[!ht]
    \centering
    \begin{subfigure}[h]{0.47\textwidth}
        \centering
        \includegraphics[width=\textwidth]{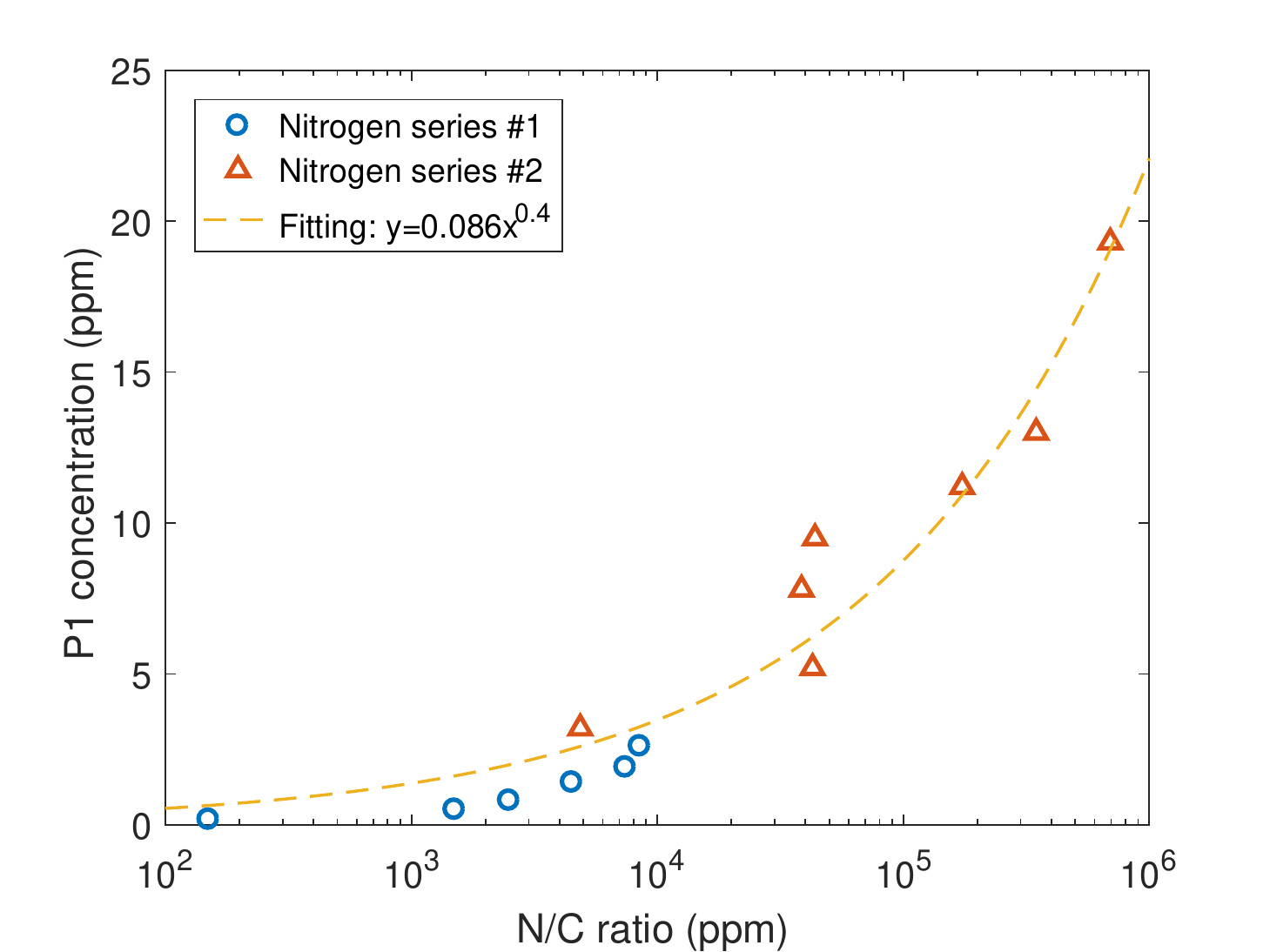}
        \put(-237,170){a)}
    \end{subfigure}
    \hfill
    \begin{subfigure}[h]{0.47\textwidth}
        \centering
        \includegraphics[width=\textwidth]{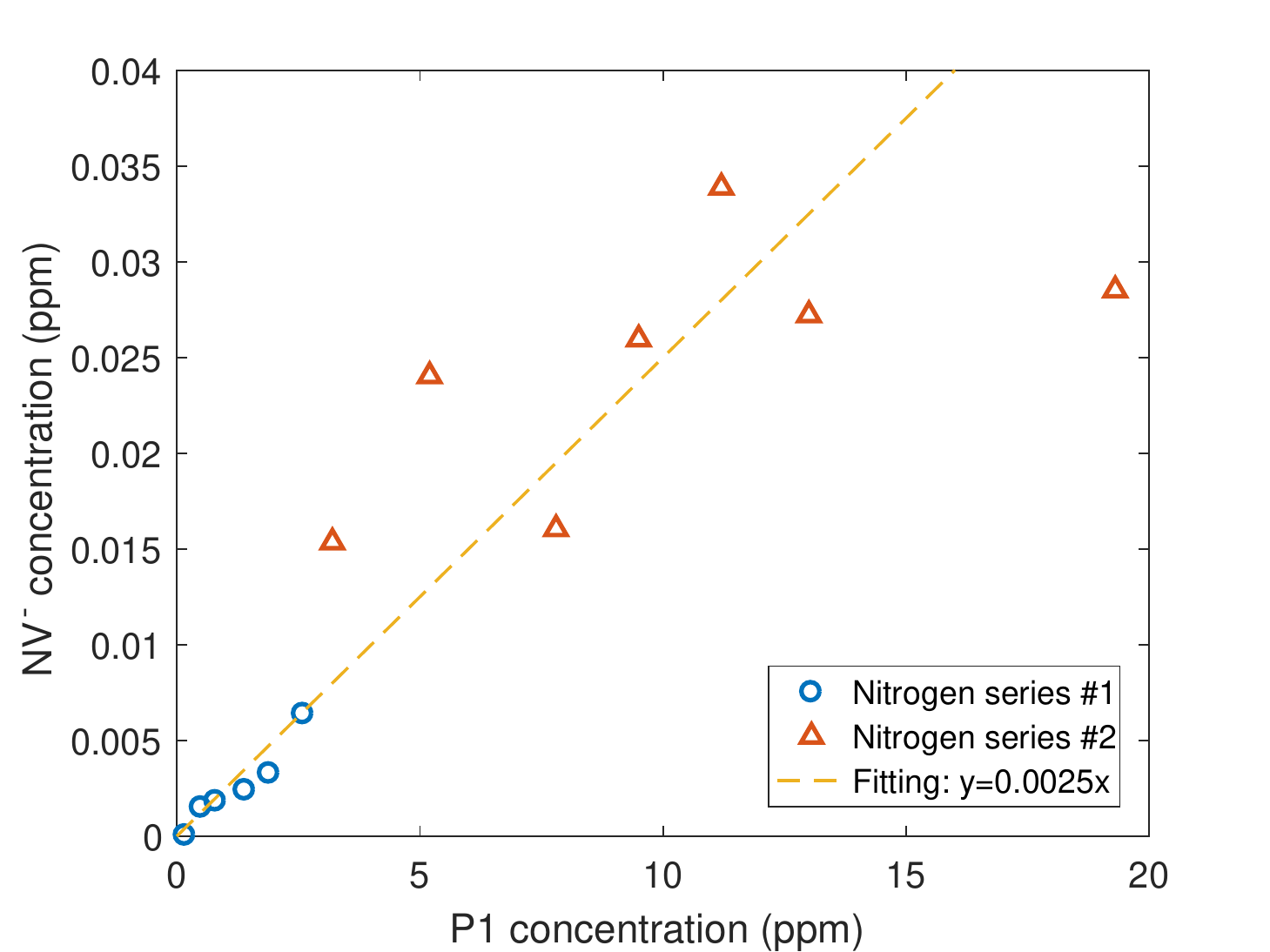}
        \put(-245,170){b)}
    \end{subfigure}
    
    \begin{subfigure}[h]{0.47\textwidth}
        \centering
        \includegraphics[width=\textwidth]{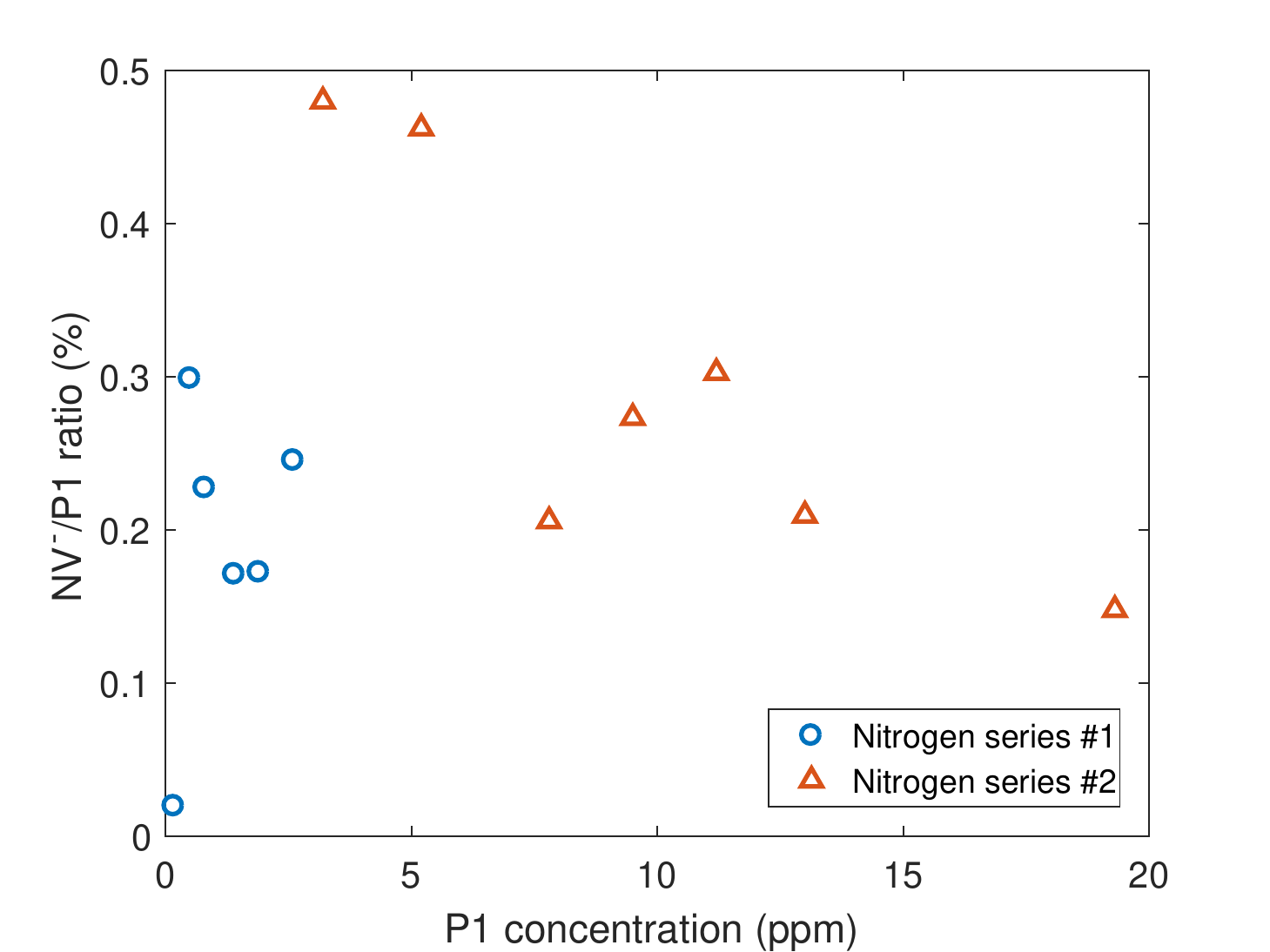}
        \put(-237,170){c)}
    \end{subfigure}
    \hfill
    \begin{subfigure}[h]{0.47\textwidth}
        \centering
        \includegraphics[width=\textwidth]{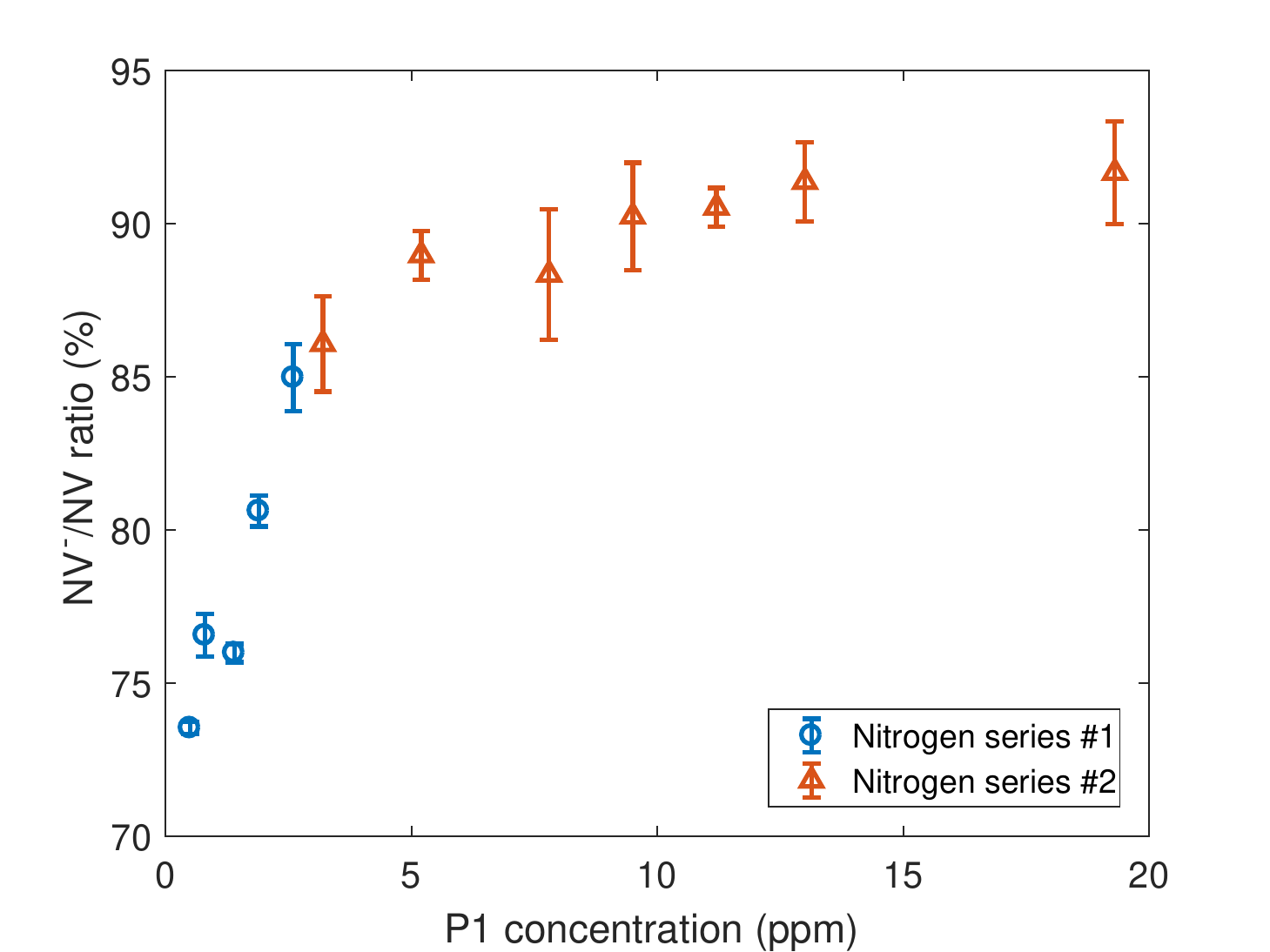}
        \put(-245,170){d)}
    \end{subfigure}
    \caption{Two nitrogen series were grown with varying N/C ratio, respectively Nitrogen series~$\#$1 with P1 of 0.2-2.6~ppm, and Nitrogen series~$\#$2 with P1 of 2.6-20~ppm. (a) The as-grown P1 concentration is positively but sublinearly correlated to the N/C flow ratio in the plasma. (b),(c) The as-grown NV$^-$ concentration is proportional to the P1 concentration, with an NV/P1 ratio of $\sim$0.25$\%$. (d) Both NV$^-$ and NV$^0$ were formed during the growth, the NV$^-$/NV ratio increases slightly with the P1 concentration. The NV-PL has a high homogeneity in the measured area, and its high precision for relative concentration measurements leads to a negligible statistical error comparing to the systematic error of $\sim$30$\%$, which was induced by the calibration from the absorption cross-section (Sec.~\ref{chap:method}), therefore the error bars have not been implemented for NV concentrations. For the same reason the P1 concentration measured by EPR with $\sim$6\% error is also not implemented. The error for the NV$^-$/NV ratio was given by the standard deviation of data acquired at different sample positions.}
    \label{fig:P1NV_grown}
\end{figure*}

We have grown the two nitrogen series with an varying N/C ratio altered by the N$_2$ to CH$_4$ flow ratio in the plasma. The N/C ratio was varied from 150 to 10$^6$ ppm during growth, resulting in an incorporated P1 concentration from 0.2 to 20 ppm (Fig.~\ref{fig:P1NV_grown}a).
The Nitrogen series $\#$1 was grown on HPHT type IIa substrates (New Diamond Technology), and the Nitrogen series $\#$2 was grown on CVD substrates (Element Six).
The growth conditions were consistent within the series with negligible variations (see Sec.~\ref{chap:samples}), with only the intended variation of the N/C ratio by increasing nitrogen flow.
We found that the P1 concentration is positively correlated to the N/C ratio, with roughly P1$\sim0.09\sqrt{ \text{N/C}}$, showing a clearly sublinear scaling, meaning more nitrogen in the reacting space leads to worse incorporation of single nitrogen atoms into the diamond lattice (note the logarithmic scale along the x axis of Fig.~\ref{fig:P1NV_grown}a).

The as-grown NV$^-$ concentration is proportional to the P1 concentration in the entire range (Fig.~\ref{fig:P1NV_grown}b), with the highest concentration of 33.9~parts per billion (ppb). This is much higher than normal as-grown HPHT diamonds, in which NV centers are largely dissociated under their growth conditions.
For both nitrogen series, the NV$^-$/P1 ratio is constant for different P1 concentrations, around 0.25\% (Fig.~\ref{fig:P1NV_grown}c), which is in the same order as previously reported values ($\sim$0.1$\%$ by De{\'a}k at el.~\cite{deak2014formation} and $\sim$0.5$\%$ by Edmonds at el.~\cite{edmonds2021characterisation}).
Based on this fact, for the same growth protocols, the relative P1 density can be roughly estimated from their as grown NV concentration, i.e. from their fluorescence intensity. 
However, we point out that a change in sample holder size and corresponding changes in the electric field distribution inside the reactor can influence the NV/P1 ratio.~\cite{Langer2022}

After growth, the NV$^-$ dominates with a NV$^-$/NV ratio above 70\% for most of the samples (Fig.~\ref{fig:P1NV_grown}d).
The low NV/P1 ratio shows that more P1 centers than vacancies were formed and NV creation in-growth is vacancy-limited. 
The large number of P1 centers provide electrons to charge the NV$^-$ center, which is consistent with the domination of NV$^-$ and makes the charge state mostly independent of the P1 centre density:
with the P1 concentration varying over two orders of magnitude, the NV$^-$/NV ratio has only slight changes. 
Nevertheless, the NV$^-$/NV ratio still shows a clear increasing trend with increasing P1 concentration, indicating that the mean distance between P1 and NV centers is also relevant for the NV charge state distribution, not only for the NV/P1 ratio.
Details of P1 and NV creations see Table.~\ref{table:NV_T2}.
The values we show are reactor- and protocol-dependent, which can be further optimized by modifying the growth conditions, but the trends reveal the general rule.

\subsubsection{Coherence time as grown}
\label{chap:T2grown}
For NV-ensemble-based magnetometry, the shot-noise limited sensitivity $\eta$ is determined by the coherence time and the NV density~\cite{rondin2014magnetometry,osterkamp2019engineering,capelliinvestigation}:
\begin{equation}
    \eta\propto\frac{1}{C\sqrt{I\cdot\tau}}
    \label{eq:sensitivity}
\end{equation}
where $C$ is the measurement contrast, $I$ is the PL intensity of NV centers, which is magnified by the NV concentration. The coherence time $\tau$ can be set to either $T_2$ or $T_2^*$ for AC or DC magnetometry respectively. A large $\sqrt{I\cdot\tau}$ leads to better sensitivity, thus both a high NV fluorescence and a long coherence time are preferred.

\begin{figure}[h!]
    \centering
    \begin{subfigure}[h]{0.97\linewidth}
        \centering
        \includegraphics[width=\linewidth]{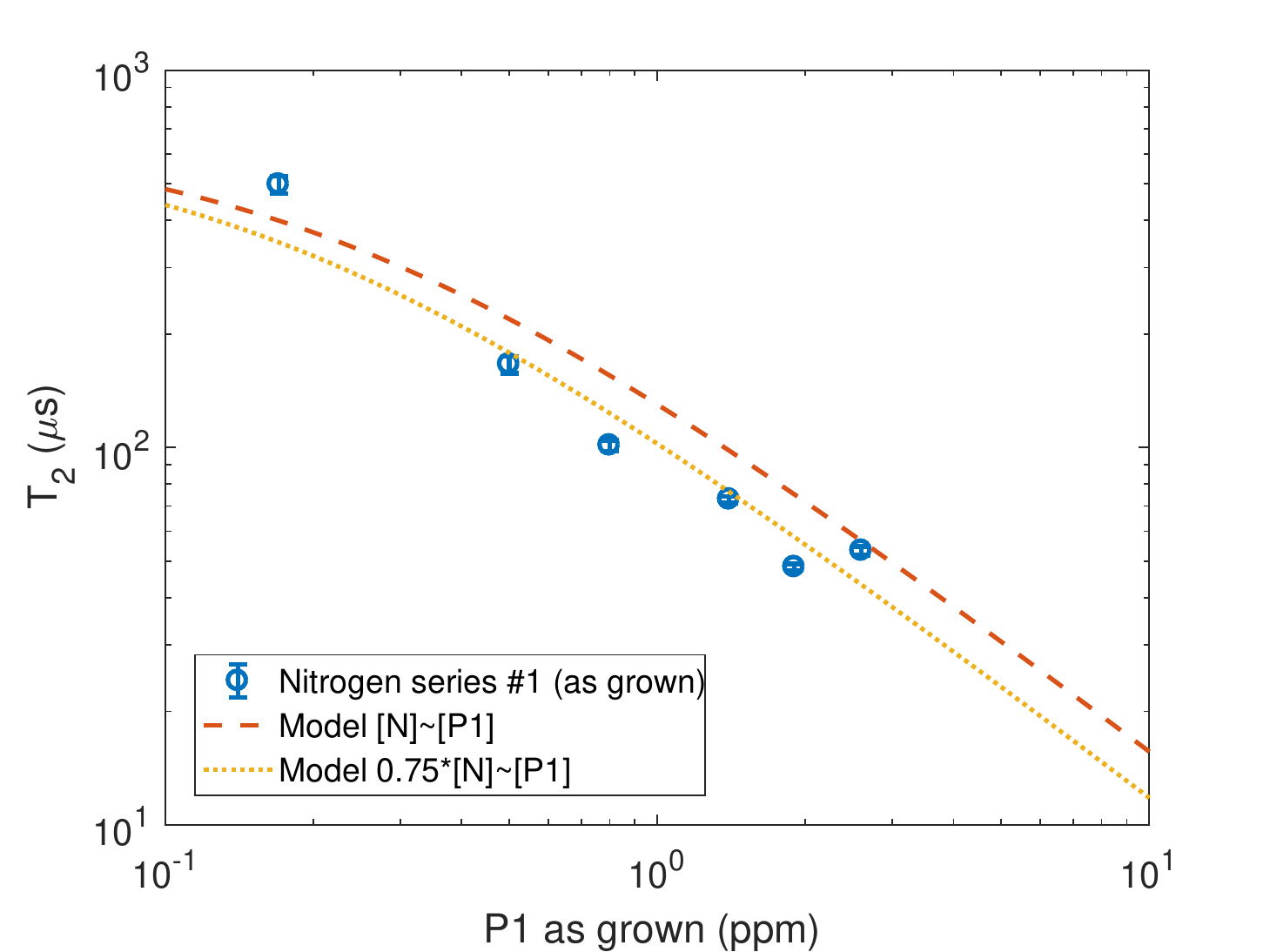}
        \put(-240,170){a)}
    \end{subfigure}
    \hfill
    \begin{subfigure}[h]{0.97\linewidth}
        \centering
        \includegraphics[width=\linewidth]{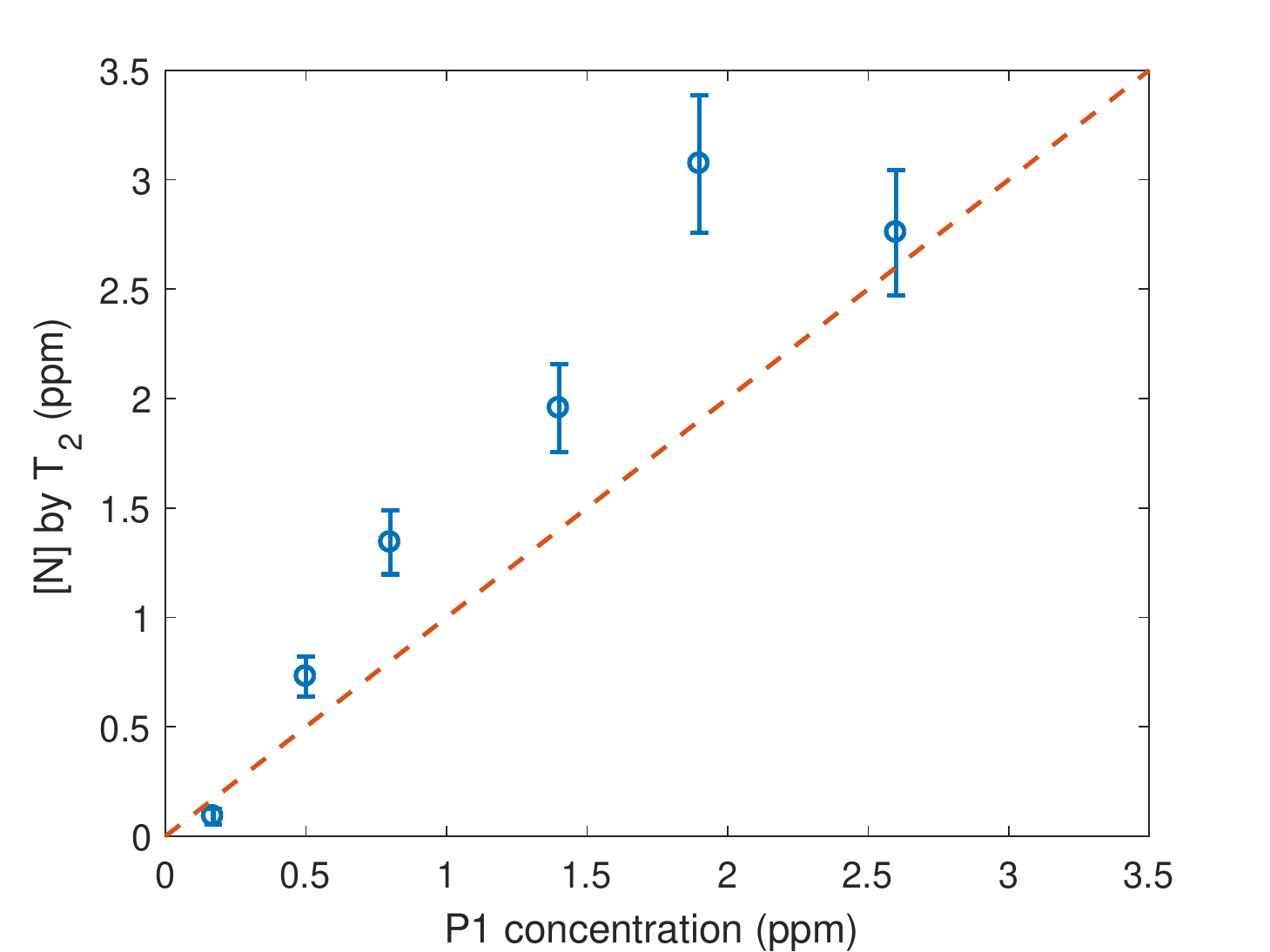}
        \put(-240,170){b)}
    \end{subfigure}
    \caption{The coherence time $T_2$ of the Nitrogen series $\#$1 measured by Hahn-echo (a) is limited by the nitrogen concentration, consistent with the theoretical model in Eq.~\ref{eq:T2-N}. Since other nitrogen-related defects (e.g. NVH centers) are often observed in as-grown CVD diamonds, but we have only measured P1 concentration for the series, we have assumed the extreme case that all nitrogen is in P1 form for 'Model [N]$\sim$[P1]', and 75$\%$ nitrogen are in P1 form for 'Model 0.75$\*$[N]$\sim$[P1]'.  (b) The nitrogen concentration calculated from $T_2$ is in scale with the directly measured P1 concentration by EPR. The error of $T_2$ is given by the fitting error (95\% confidence interval).}
    \label{fig:T2}
\end{figure}

Our nitrogen series showed significant NV fluorescence after growth, which correlated positively to the P1 concentration. The Hahn-echo coherence time $T_2$, however, is shorter for higher P1 concentration (blue circles in Fig.~\ref{fig:T2}a): 498 $\mu$s for 0.2 ppm P1, while only 53 $\mu$s for 2.6 ppm P1 (details see Table.~\ref{table:NV_T2}).

The decoherence time T$_2$ is set by fluctuating magnetic dipoles surrounding and coupling to the NV centres. It has been suggested \cite{hanson2006room, bauch2020decoherence} that nitrogen electrons are the major source of decoherence when they are above a concentration of $\sim$0.5~ppm. For lower concentrations typically the nuclear spins of $^{13}$C are the dominating magnetic dipole noise limiting the coherence time. Bauch~{\em et al.}~\cite{bauch2020decoherence} has suggested a model linking nitrogen concentration and coherence time as:
\begin{equation}
    \frac{1}{T_2([N])} = B_{NV-N}\cdot[N]+\frac{1}{T_{2,other}}
    \label{eq:T2-N}
\end{equation}
where $B_{N-NV}=2\pi\times(1.0\pm0.1)$ kHz/ppm is the nitrogen-dominated NV decoherence rate, $[N]$ is the nitrogen concentration, and $T_{2,other}=694\pm82$ $\mu$s comes from the decoherence factor independent of nitrogen. 
In this model, they determined the total nitrogen concentration in the crystal by the secondary ion mass spectroscopy (SIMS), while we measured the single substitutional nitrogen (P1 centers) via EPR. 
In Fig.~\ref{fig:T2}a, the two curves show the model predicting the coherence times for P1 concentrations constituting a 100$\%$ of $[N]$ (red dashes) and for 75$\%$ of it (orange dotted).
As we observed other nitrogen-related defects (especially NVH centers) in the EPR experiment for our nitrogen series, we would expect a $[P1]/[N]$ ratio lower than 100$\%$.
For the curve that 75$\%$~[N] being P1 centers, the coherence time T$_2$ of our nitrogen series showed high consistency with the model, indicating that 75$\%$ can be the estimated $[P1]/[N]$ ratio for the series. 
However, as we noticed that more NVH centers were formed in our Nitrogen series $\#$2 than $\#$1, we think that the $[P1]/[N]$ ratio is varying for different nitrogen-doping levels--it can be smaller for higher nitrogen concentrations. 
This can potentially explain the long T$_2$ of the lowest P1 sample, which is much longer than the theoretical value given by the 0.75[N]$\sim$[P1] model. 
Fig.~\ref{fig:T2}b illustrates the correlation between measured [P1] and [N] calculated from $T_2$ according to Eq.~\ref{eq:T2-N}, further indicating that other nitrogen-related defects exist in the series, but with minor influences.

The demand for high NV fluorescence and long coherence time is contrary, calling for a balance between nitrogen incorporation and coherence time. For this purpose, the nitrogen-doped CVD growth performs competitive as it is highly controllable to obtain prospective combinations of these two factors consistently.
We have proven that after growth, nitrogen-doped CVD diamonds can acquire sufficient NV$^-$ concentrations, as well as long coherence times for sensing requirements. 
The sensitivity can be further improved by irradiation and annealing steps after growth, which convert P1 centers into NV centers efficiently, increasing the $NV/P1$ ratio while not shortening the $T_2$ (Eq.~\ref{eq:T2-N}). According to Eq.~\ref{eq:sensitivity}, with an enhanced NV PL signal $I$ and a consistent coherence time $\tau$, the sensitivity $\eta$ can be improved. Optimizing the irradiation we can maximize the P1-to-NV conversion rate, while performing the optimal combination of high NV PL signal, charge stability, and long $T_2$ time.
\subsection{NV-ensembles after e-beam irradiation}
To efficiently enhance the NV concentration in the diamond, subsequent irradiation and annealing treatments are commonly conducted, which have been discussed in previous works~\cite{rubinas2021optimization,edmonds2021characterisation}.
Multiple irradiation types are commonly applied for vacancy creation in diamonds, including irradiation with electrons, ions, high-energy photons or neutrons~\cite{ashbaugh1988gemstone,mita1996change,campbell2000radiation}. 
Electron irradiation creates mainly isolated vacancies homogeneously~\cite{barry2020sensitivity}, while avoiding crystal damages, which is ideal for NV ensembles.
Moreover, the electron irradiation can realize a penetration depth up to millimeters~\cite{campbell2000radiation}, which ensures the vacancy creation through the bulk diamond plate.
The in-situ or subsequent annealing then mobilises vacancies in the diamond~\cite{davies1976optical,collins2009annealing,deak2014formation,capelli2019increased}, which are trapped by P1 centers to form NV centers~\cite{acosta2009diamonds,davies1977charge,orwa2011engineering}.

We applied e-beam irradiation and subsequent annealing to further create NV centers in our nitrogen-doped CVD diamonds, enhancing the NV/P1 ratio, while not expecting a shortening in coherence time $T_2$, since it is limited by the total nitrogen concentration in the diamond,  which should be consistent. 
In this sense, the irradiation condition should be optimized to maximize the P1 to NV$^-$ conversion rate: when a P1 center and a vacancy combine into an NV center, it then needs other P1 centers as the donor to provide an electron to charge NV$^-$~\cite{collins2002fermi,fu2010conversion,haque2017overview,hauf2011chemical}. Consequently, when large amounts of P1 centers are converted to NV centers, an insufficient remaining P1 concentration potentially leads to more NV$^0$ formation rather than NV$^-$. The NV$^0$ center is detrimental for two reasons: firstly it cannot be used for sensing and thus contributes as a fluorescence background that weakens the ODMR contrast~\cite{edmonds2021characterisation}, thus reduces the sensitivity (Eq.~\ref{eq:sensitivity}); secondly, more NV$^0$ indicates a lack of electron availability, which leads to stronger photo-ionisation~\cite{jeske2017stimulated}, i.e. more NV$^-$ is converted into NV$^0$ when pumped with green light. 
In conclusion, high NV$^-$/P1 and NV$^-$/NV ratios are the two key factors while optimizing irradiation. 

The irradiation fluence influences the vacancy concentration linearly~\cite{campbell2000radiation}, consequently, diamonds with different P1 concentrations call for different optimal irradiation fluences. 
We investigated this optimum for diamonds with a fixed P1 density, by irradiating them with varying irradiation fluences and energies. 
Throughout studies of their NV concentrations, NV charge state distributions and defect transformations, we suggested general rules to determine optimal irradiation fluences for different P1 densities.
Furthermore, by optimizing the irradiation condition, we can combine enhanced NV$^-$ concentrations with long coherence times to achieve improved sensitivities.

\subsubsection{Optimizing irradiation fluence for NV creation}
\label{chap:fluence}
\begin{figure*}[!ht]
    \begin{subfigure}[h]{0.47\textwidth}
        \centering
        \includegraphics[width=\textwidth]{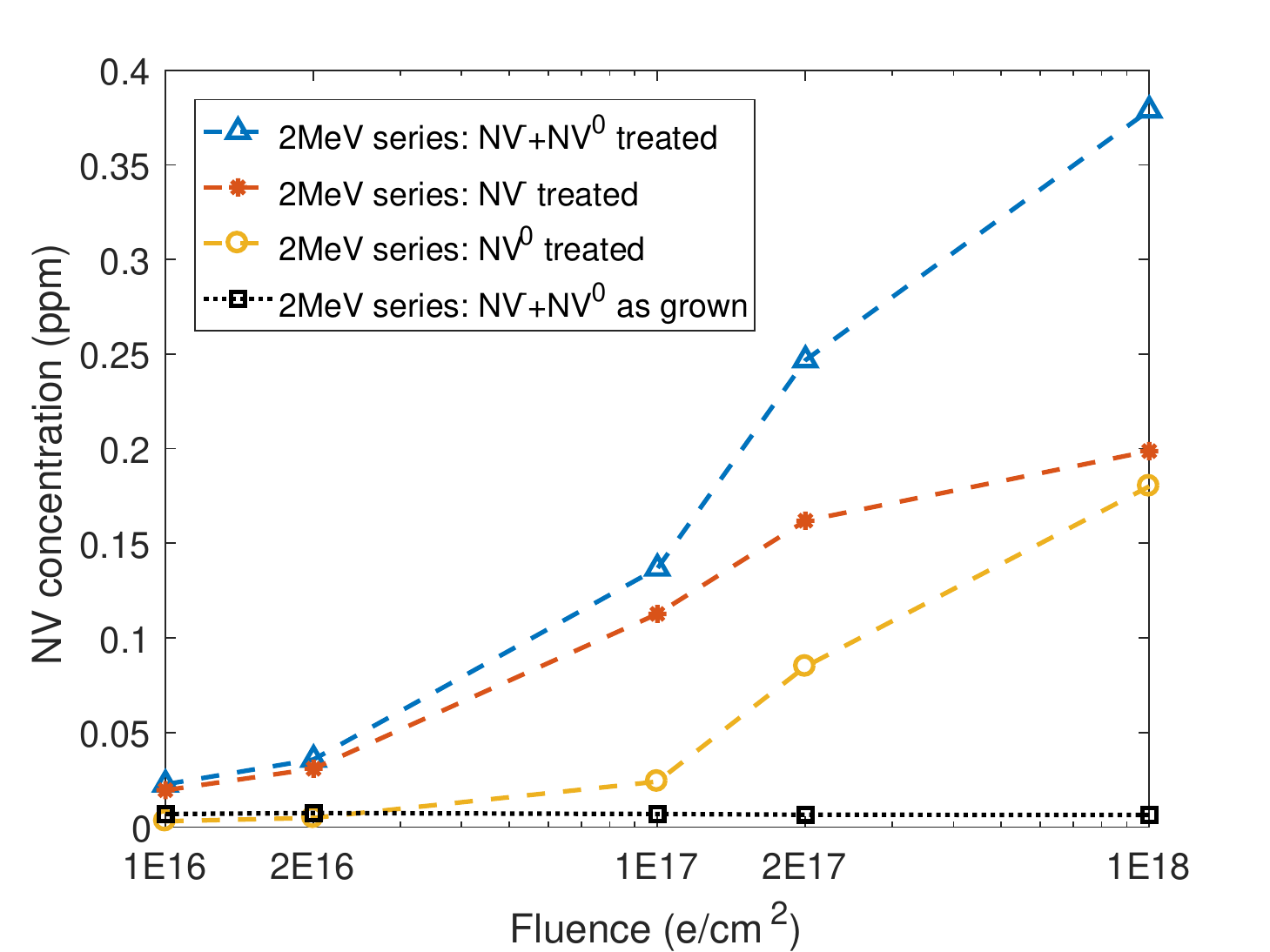}
        \put(-237,170){a)}
    \end{subfigure}
    \hfill
    \begin{subfigure}[h]{0.47\textwidth}
        \centering
        \includegraphics[width=\textwidth]{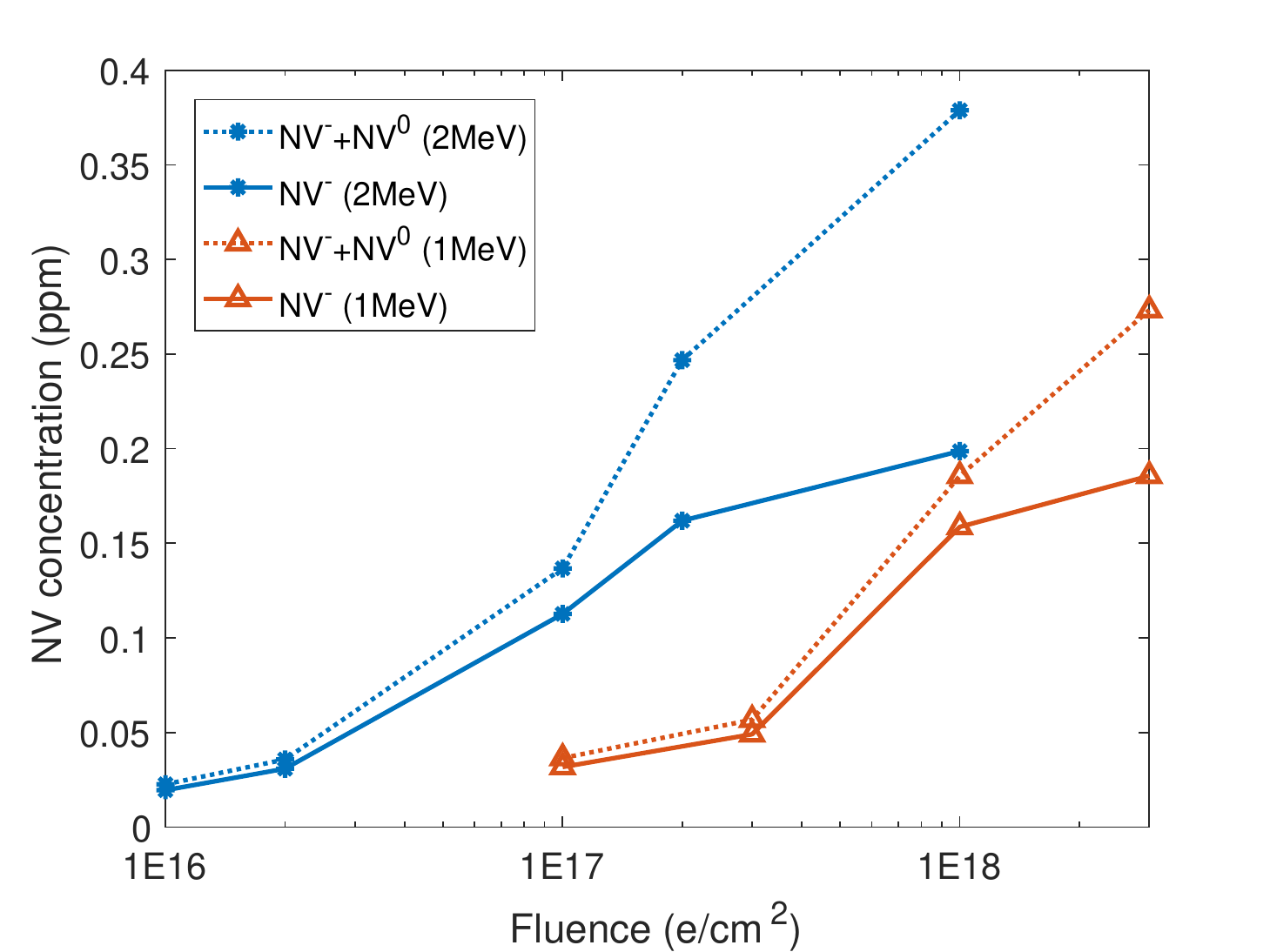}
        \put(-237,170){b)}
    \end{subfigure}
    
    \begin{subfigure}[h]{0.47\textwidth}
        \centering
        \includegraphics[width=\textwidth]{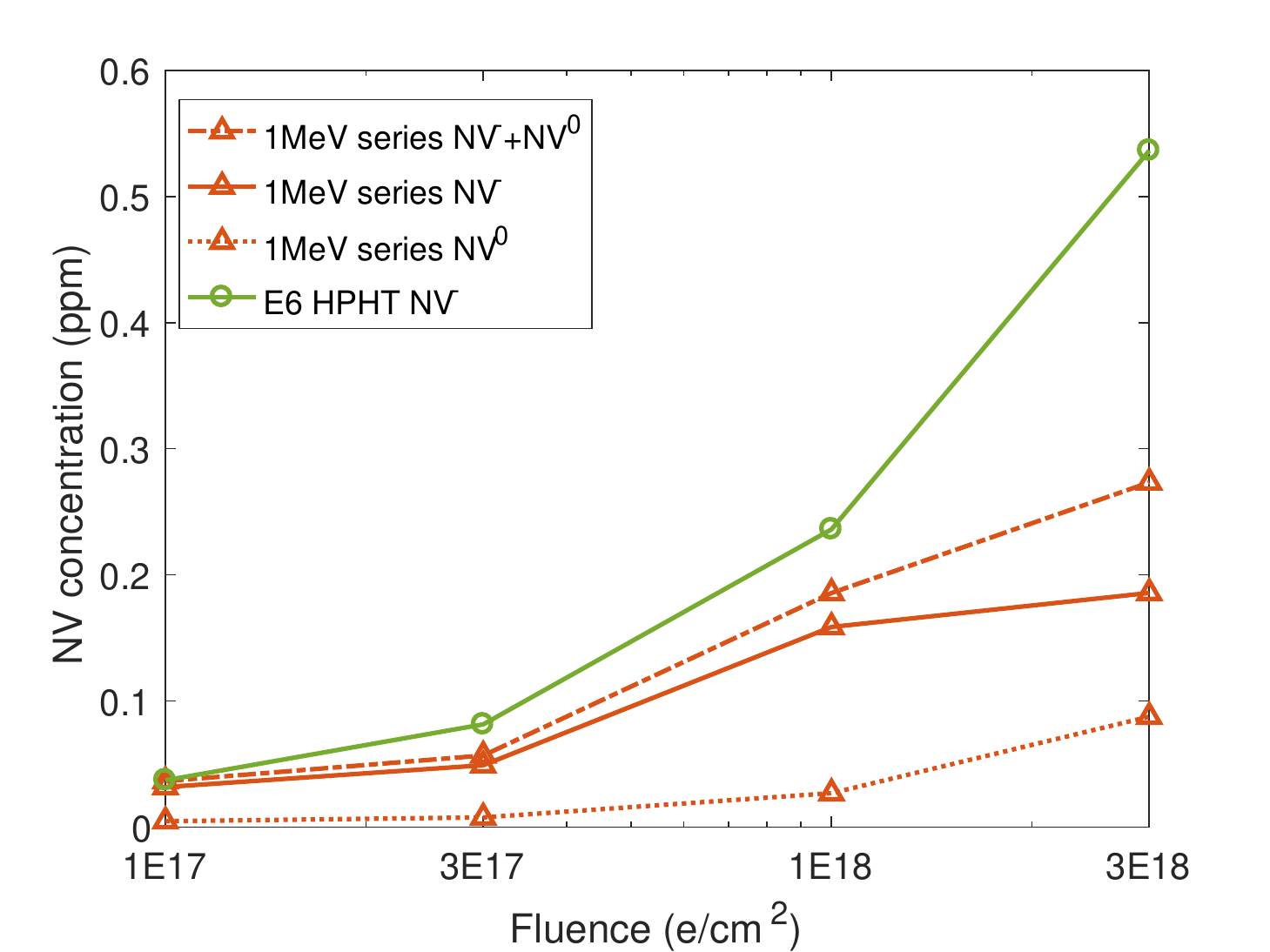}
        \put(-237,170){c)}
    \end{subfigure}
    \hfill
    \begin{subfigure}[h]{0.47\textwidth}
        \centering
        \includegraphics[width=\textwidth]{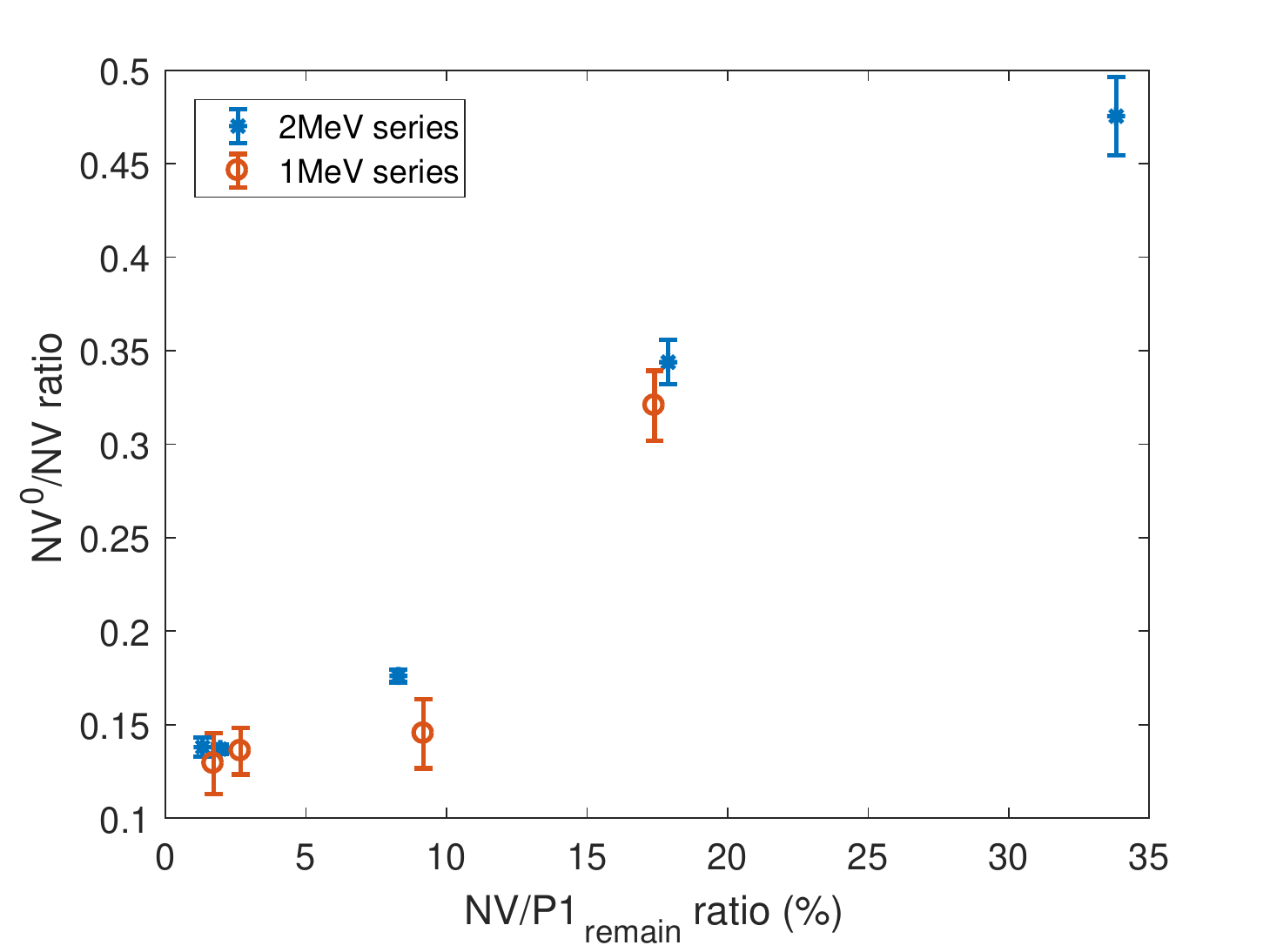}
        \put(-237,170){d)}
    \end{subfigure}
    \caption{Two CVD series with P1$\sim$2.2~ppm have been e-beam irradiated with respectively 2~MeV and 1~MeV, with varying fluences. (a) For the 2~MeV series, the NV concentration increased with the irradiation fluence with the optimal fluence between 1E17-2E17~e/cm$^2$, with a high NV concentration and high NV$^-$/NV ratio. (b) The 1~MeV series shows similar behavior as the 2~MeV series, but shifts towards higher irradiation fluence by an order of magnitude. (c) Under the same irradiation and annealing conditions, HPHT diamonds with P1$>$20~ppm show higher NV concentration (with only NV$^-$) than CVD diamonds with P1$\sim$2.2~ppm. (d) NV$^0$/NV ratio is positively correlated to the NV/P1$_{remain}$ ratio in CVD diamonds, when NV/P1$_{remain}>$10$\%$, the NV charge state shifts significantly towards NV$^0$. Errors are defined as in Fig~\ref{fig:P1NV_grown}.}
    \label{fig:Irr}
\end{figure*}

To optimize the e-beam irradiation, and to understand how the fluence and energy influence the NV creation, we have grown the two irradiation series with fixed growth parameters, which showed high PL consistency after growth (Fig.~\ref{fig:Irr}a). The P1 concentration of the series, estimated from the UV-Vis spectrum, is also consistent with $\sim$2.2~ppm. 
The two series have been irradiated with respectively 2~MeV and 1~MeV electrons at the room temperature, with varying fluences (Fig.~\ref{fig:Irr}). The annealing was performed after irradiation 1000~$^{\circ}$C for 2~h in vacuum (Sec.~\ref{chap:samples}).
Additionally, we irradiated four HPHT diamonds (Element Six, P1$>$20~ppm) with 1~MeV electrons, as an extension of our study for higher P1 concentrations.

after treatment, for both the 2~MeV and 1~MeV series, the total NV concentration increased significantly, positively correlated to the irradiation fluence, with strong variations on the NV charge state distribution (Fig.~\ref{fig:Irr}). 
For the 2~MeV series, we can see the expected transition starting from 1E17~e/cm$^2$ where NV$^-$ creation saturates and mostly NV$^0$ is produced with increasing fluence (Fig.~\ref{fig:Irr}a). 
Around 1E17-2E17~e/cm$^2$ seems to be the optimal point with still enough P1 centers in the diamond to charge NV centers, but no excess which is positive for longer coherence times. 
The highest P1 to NV- conversion rate in this series was $\sim$8.9\% for 1E18 e/cm$^2$, however a poor NV$^-$/NV ratio of only 53$\%$ was achieved. For 2E17~e/cm$^2$ the P1 to NV- ratio was $\sim$7.3\%, maintaining an NV$^-$/NV ratio of 66$\%$, indicating better robustness against photo-ionization.

The 1~MeV series showed similar behavior in NV concentration and charge states, but shifted towards higher fluence by an order of magnitude (Fig.~\ref{fig:Irr}b), the optimum then shifted to a range in 1-3E18~e/cm$^2$. 
This strong influence of the electron energy is in contrast to theoretical calculations~\cite{campbell2000radiation} which predict that the  2MeV irradiation with 1E17~e/cm$^2$ creates roughly 1.1 ppm vacancies, while for 1~MeV it is 0.9~ppm with the same fluence, the vacancy concentrations differ slightly but not by an order of magnitude. 
Experimentally, our results clearly contradict this and show that the question of the influence of electron energy on vacancy creation needs to be reconsidered. 
However we point out that the two different energies were performed at different facilities and suggest studies of varying energies in the same facility for future studies.

For the same irradiation conditions, the NV concentration also depends on the initial P1 concentration: as in Fig.~\ref{fig:Irr}c, HPHT Ib diamonds showed higher NV concentration (with only NV$^-$) than the CVD series, since they contained more initial P1 centers before irradiation. This indicates that the vacancies created via irradiation were not fully converted into NV centers in the CVD series. However, the deviation between the HPHT series and the CVD series is small at low fluences and only becomes significant when the NV creation reaches a point where it is limited by the overall P1 availability.

In Sec.~\ref{chap:NV_asgrown} we have discussed vacancy-limited NV charge state distribution after CVD growth, as much fewer vacancies than P1 centers were in the diamond. 
After irradiation, we went to the point that the vacancy concentration (up to $\sim$30~ppm for 3E18~e/cm$^2$ with 1MeV) was in the same order with--or even higher than--the P1 concentration (2.2~ppm), vacancies have not been fully converted after annealing and the NV creation was P1-limited. 
Since NV$^-$ centers are charged by P1 centers, we suspected that the NV$^-$/NV ratio is dependent on the remaining P1 centers after conversion to charge the NV centers. 
As support, for the two CVD irradiation series, we measured the P1 concentration twice, as after growth $[P1_{grown}]$ and after irradiation and annealing $[P1_{remain}]$. 
After treatment, the NV$^-$/NV ratio seems in fact to be determined by the ratio $R_{re}$=NV/P1$_{remain}$, independent of irradiation fluence or energy (Fig.~\ref{fig:Irr}d). 
When $R_{re}>$10$\%$, the NV$^0$ ratio starts to increase largely.
As P1$_{grown}$ is partially converted to NV centers (and possibly other defects) during the treatment, $[P1_{remain}]$ as the final state is always less than $[P1_{grown}]$. 
The conversion rate from P1$_{grown}$ to NV centers after treatment, denoted as $R_{con}$=NV/P1$_{grown}$, always fulfills $R_{con}<R_{re}$, consequently $R_{con}<$10$\%$ is required to achieve NV$^0$/NV$<$20$\%$.
Given that the photo-ionization can be largely promoted by higher green laser power, the NV$^0$/NV ratios we measured here with lower laser power (10~$\mu$W confocal) represent the intrinsic NV-charge-state distribution and usually get worse at high power densities. 
Therefore to achieve a high charge state stability, we take $R_{con}<$10$\%$ as an important criteria for the fabrication.

Another criteria that might be interesting for some applications is $R_{re}\sim$35$\%$, for which half of NV centers will be in the neutral form NV$^0$. This means to remain NV$^-$ dominated, $R_{re}<$~35$\%$ is required.
For an initial P1 concentration of 2.2~ppm, the optimum can be set depending on different purposes. 
When focusing on the NV-charge-state stability, we would suggest an optimum of 1E17~e/cm$^2$ with 2MeV electron, or 1E18~e/cm$^2$ with 1MeV electron, resulting an NV$^-$/NV ratio of $\sim$82-86$\%$.
When focusing on the total NV creation, the optimum can be set to 2E17~e/cm$^2$ with 2MeV electron, or 3E18~e/cm$^2$ with 1MeV electron, resulting an P1$_{grown}$ to NV$^-$ conversion rate $R^-_{con}$ of $\sim$7.3-8.4$\%$ while remaining an NV$^-$/NV ratio of $\sim$66-68$\%$. (More details see Table.~\ref{table:Irr})

We have also noticed that previous works have applied different methods to determine the NV concentration.
PL measurements are often applied with different calibration methods, for example comparing the NV-PL intensity with a single NV center~\cite{stanwix2010coherence,osterkamp2019engineering}, or calibrating the NV-PL with the absorption cross-section (as applied in this work)~\cite{capelli2019increased}. 
The UV-Vis absorption spectrum at cryogenic temperature has also been used to determine NV$^-$ and NV$^0$ concentrations respectively~\cite{edmonds2021characterisation,rubinas2021optimization}. 
EPR methods have been also applied in a few cases \cite{shames2017fluence,khan2013colour}, but often requires complementary methods for low NV concentrations. 
NV-PL based methods are most commonly used, although the NV fluorescence can be suppressed in some cases (e.g. by higher nitrogen concentration~\cite{Capelli2022} or graphitization~\cite{waldermann2007creating}), it is still an excellent way to evaluate the NV concentration, as it links directly to the sensing applications, and in this sense it plays even more important role than the 'absolute' number of NV spins.
Unfortunately, until now there is no systematic study of consistency between different methods. Consequently comparisons between results obtained by different methods should be treated carefully. 
Nevertheless the trends and rules we discussed for NV creation should be generally applicable in CVD diamonds.

\begin{table*}[h!]
\footnotesize
\centering
 \begin{tabular}{M{4em} M{4em} M{8em} M{9em} M{9em} M{9em}} 
  \toprule
  Series & Sample & Fluence (e/cm$^2$) & NV$^-$/NV ratio ($\%$) & NV/P1$_{remain}$ ($\%$) & NV$^-$/P1$_{grown}$ ($\%$) \\
  \hline
  \multirow{5}{4em}{\centering 2~MeV series}& I2-01 & 1E16 & 86.2 $\pm$ 0.5 & 1.3 & 0.9\\
  & I2-02 & 2E16 & 86.3 $\pm$ 0.2 & 1.9 & 1.4 \\ 
  & I2-04 & 1E17 & 82.4 $\pm$ 0.3 & 8.3 & 5.1 \\ 
  & I2-05 & 2E17 & 65.5 $\pm$ 1.2 & 17.6 & 7.3 \\ 
  & I2-08 & 1E18 & 52.5 $\pm$ 2.1 & 33.3 & 8.9 \\ 
  \midrule
  \multirow{4}{4em}{\centering 1~MeV series} & I1-39 & 1E17 & 87.1 $\pm$ 1.6 & 1.7 & 1.4 \\ 
  & I1-50 & 3E17 & 86.4 $\pm$ 1.3 & 2.7 & 2.2 \\ 
  & I1-28 & 1E18 & 85.5 $\pm$ 1.9 & 9.5 & 7.2 \\ 
  & I1-29 & 3E18 & 67.9 $\pm$ 1.9 & 15.3 & 8.4 \\ 
  \bottomrule
 \end{tabular}
\caption{NV charge states distribution and P1 to NV conversion rates of different irradiation energies and fluences. All samples have an initial P1 concentration of $\sim$2.2 ppm.}
\label{table:Irr}
\end{table*}

\subsubsection{Vacancy creation and defect transformations during irradiation and annealing via spectral studies}
\begin{figure}[h!]
    \centering
    \begin{subfigure}[h]{0.97\linewidth}
        \centering
        \includegraphics[width=\linewidth]{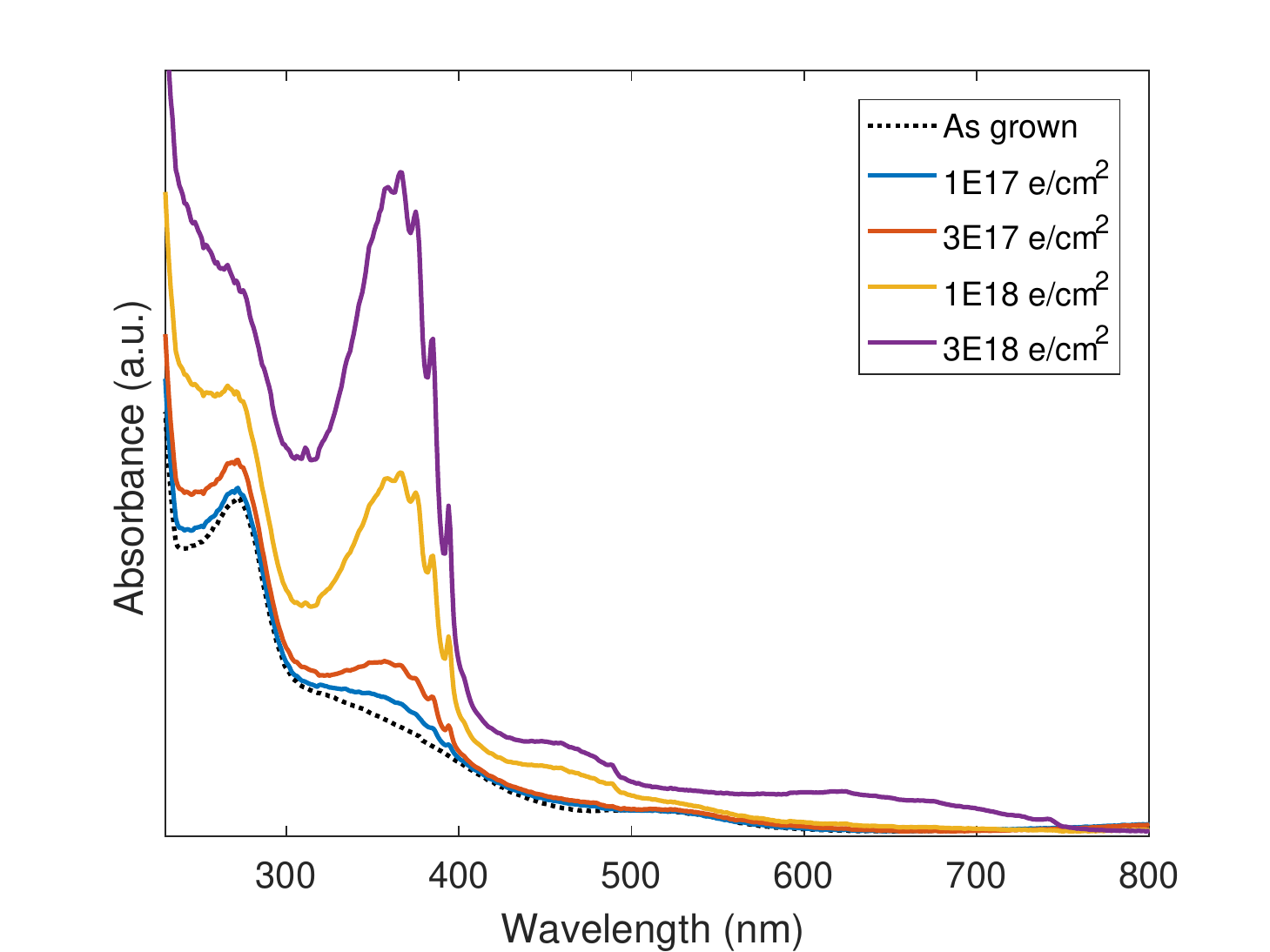}
        \put(-230,160){a)}
    \end{subfigure}

    \begin{subfigure}[h]{0.97\linewidth}
        \centering
        \includegraphics[width=\linewidth]{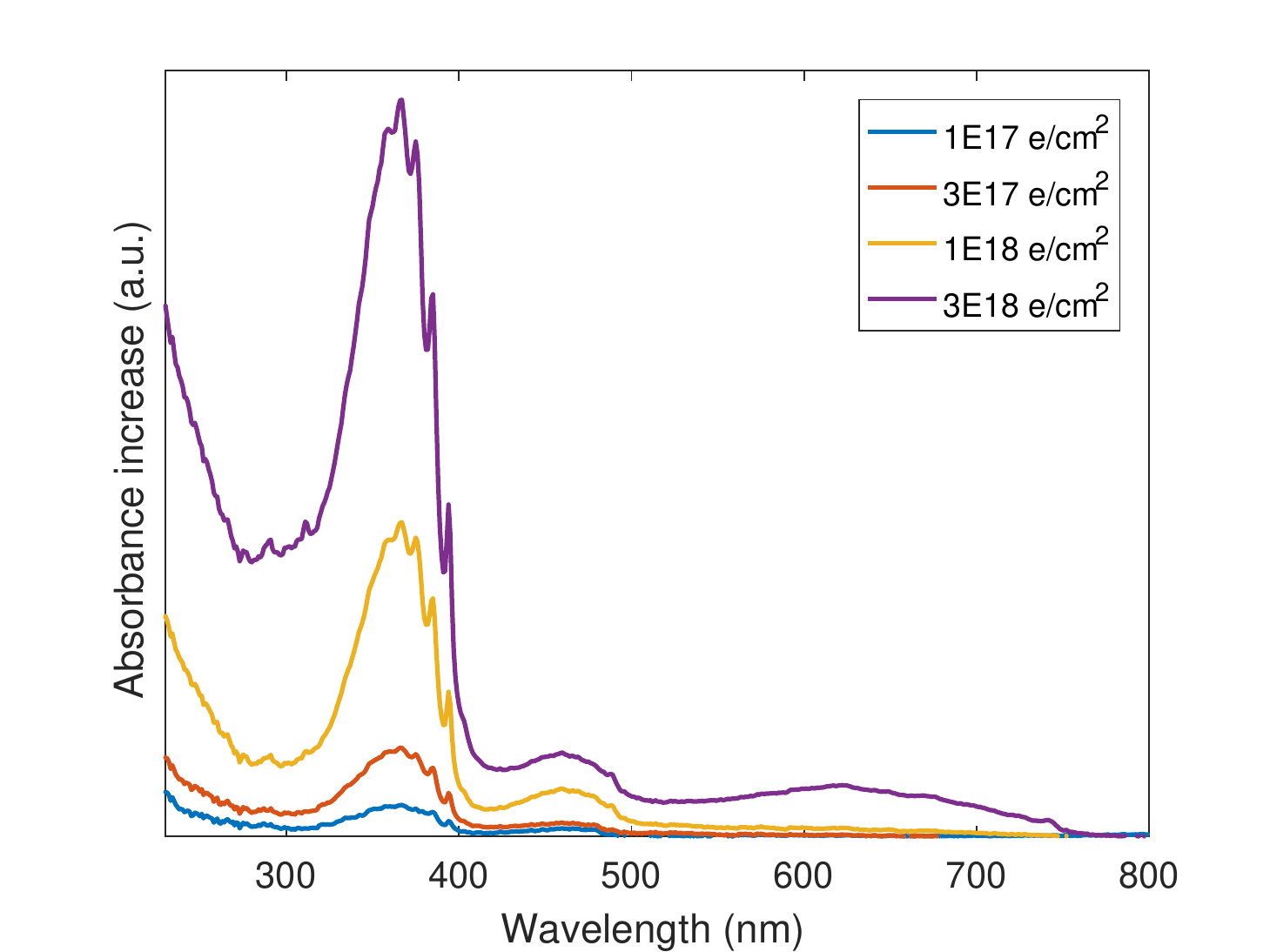}
        \put(-230,160){b)}
    \end{subfigure}
    \caption{(a) Absorption spectra and (b) absorption increases (subtracted by the as-grown spectrum) of the 1~MeV irradiation series after irradiation, before annealing. The spectra show clear ND1 bands (V$^-$, with a ZPL at 393~nm and tow phonon replicas at 375~nm and 384~nm), increasing with the irradiation fluence. For the highest fluence (3E18e/cm$^2$)m, the GR1 band (V$^0$, with a ZPL at 741~nm and a side band at $\sim$500-750~nm) can be observed. }
    \label{fig:Abs_Irr}
\end{figure}

Analysing the NV concentration and charge states after irradiation and annealing steps provides us an overview of the 'final state' for the NV creation. 
The spectral study gives us further insight of the vacancy creation, the vacancy charge states behavior, and how they influence the NV creation during the treatment steps.
To better understand this, and to visualise the defect transformations via irradiation and annealing, we study the absorption spectrum of the 1~MeV series in the UV-Visible range. 
Fig.~\ref{fig:Abs_Irr}a shows the absorption spectra after irradiation with different fluences. 
Subtracting the irradiated spectra from the as-grown ones, see Fig.~\ref{fig:Abs_Irr}b, we can clearly identify the creation of single vacancies and distinguish its negatively charged and neutral form. 
The irradiated spectra showed a significant change in a band centered at $\sim$365~nm, reported to be the ND1 band absorption band caused by negatively charged single vacancies (or V$^-$)~\cite{dyer1965irradiation,zaitsev2018defect}, with three small peaks at 393~nm, 375~nm and 384~nm, corresponding to its zero phonon line (ZPL) and two phonon replicas respectively. 
Our result demonstrated that the ND1 band increased with irradiation fluence, conforming with the argument that electrons should mainly create isolated vacancies, which accept electrons from the electron-donating P1 centers. 
For a strong irradiation fluence, for instance 3E18 e/cm$^2$ curves in Fig.~\ref{fig:Abs_Irr}, a significant GR1 band (corresponds to V$^0$) appeared, showing a ZPL at 741~nm and a broad feature from $\sim$500-750~nm ~\cite{dobrinets2016hpht}. GR1 centers are normally generated by high energy irradiation, although a high nitrogen concentration can promote the conversion from GR1 to ND1~\cite{dobrinets2016hpht,zaitsev2018defect}. 
At this stage of the processing, the vacancies are the main electron acceptors and these will be converted to NV centers in the subsequent annealing step. The main electron donors are the P1 centers. Assuming that the vacancies are charged by the P1 centers and that with increasing vacancy creation, there might be a point where the availability of electrons is limited by the P1 density, this point could be identified by the appearance of neutral vacancies (GR1 centers) instead of negatively charged vacancies (ND1 centers). As the vacancies are converted to NV centers during annealing, the appearance of neutral vacancies before annealing might indicate the appearance of neutral NV centers after annealing. This is supported by the fact that only the sample with irradiation fluence of 3E18~e/cm$^-2$ shows a strong GR1 band after irradiation and significant NV$^0$ centers after annealing, see Fig.~\ref{fig:Irr}b. 
Thus, the appearance of GR1 centers in irradiation can potentially be used to determine the ideal fluence for NV creation, even before annealing. 

\begin{figure}[h!]
    \centering
    \begin{subfigure}[h]{0.97\linewidth}
        \centering
        \includegraphics[width=\linewidth]{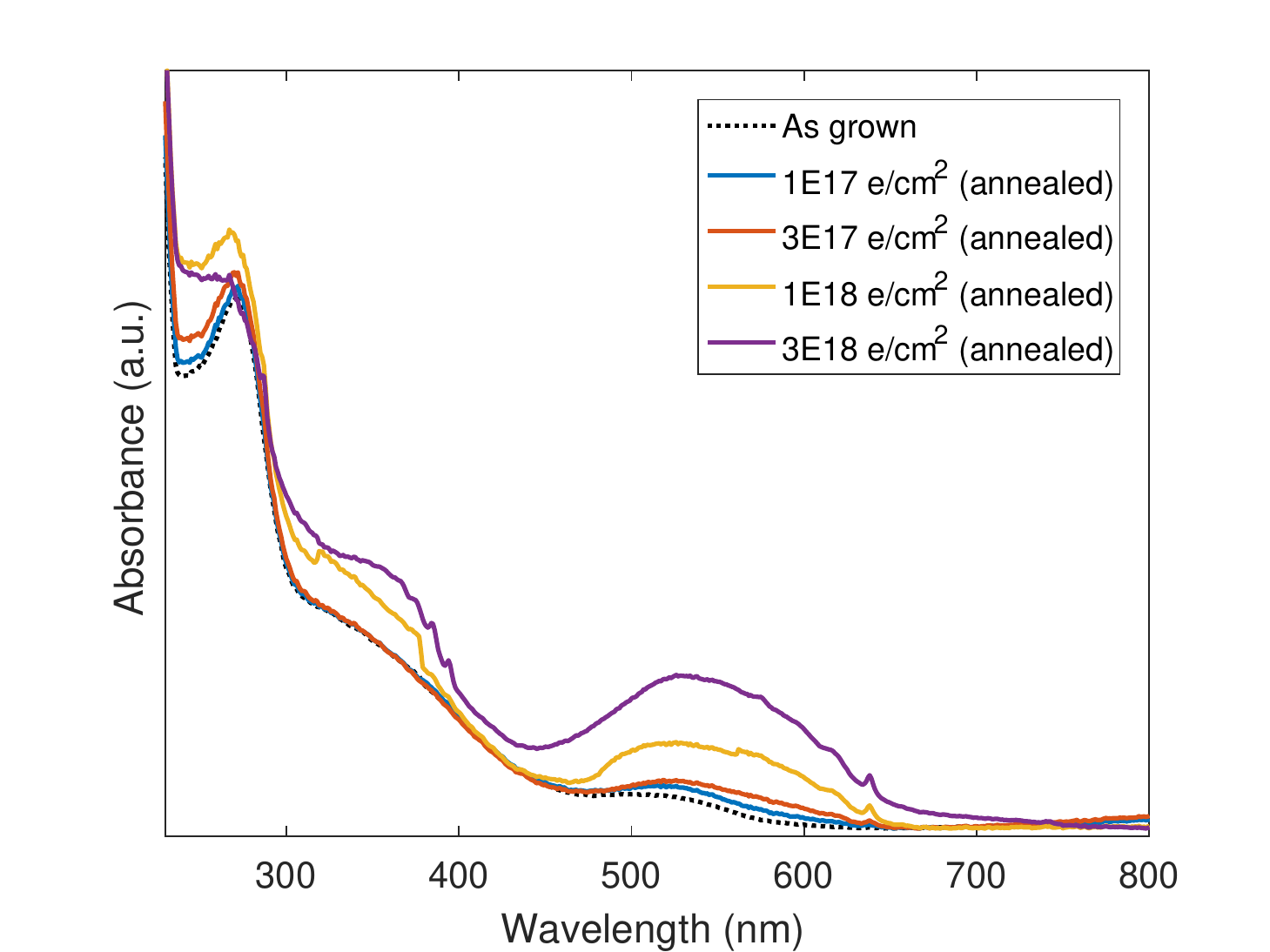}
        \put(-230,160){a)}
    \end{subfigure}

    \begin{subfigure}[h]{0.97\linewidth}
        \centering
        \includegraphics[width=\linewidth]{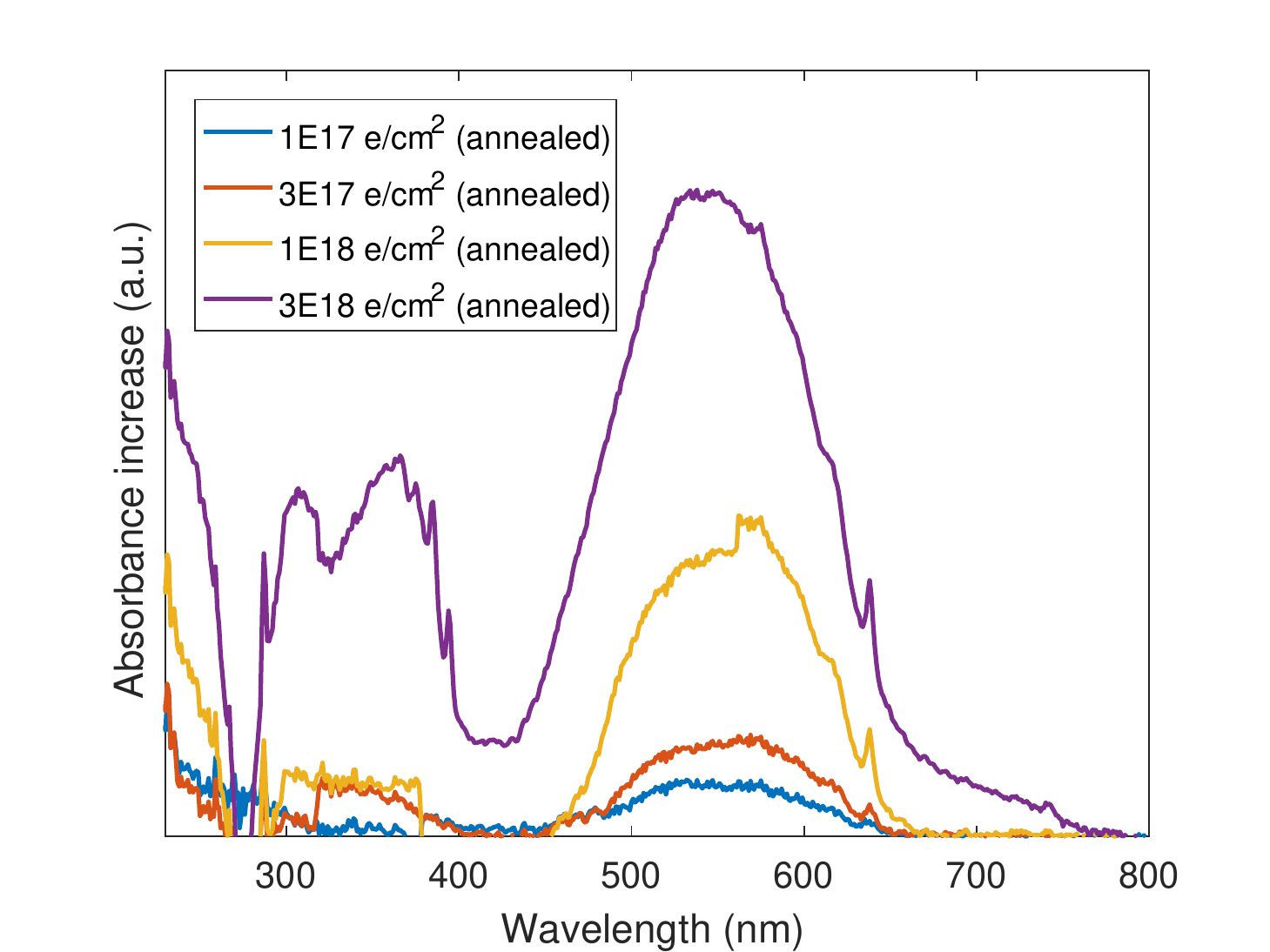}
        \put(-230,160){b)}
    \end{subfigure}
    \caption{(a) Absorption spectra and (b) absorption increases (subtracted by the as-grown spectrum) of the 1 MeV irradiation series after annealing. The spectra show significant decreases in the ND1 band and increases of the NV band. The NV concentration is proportional to the irradiation fluence. For the highest fluence, both V$^-$ and V$^0$ centers were not fully converted.}
    \label{fig:Abs_Ann}
\end{figure}

Additionally, the spectrum after irradiation showed an increase in the 'ramp', which was defined by Khan~{\em et al.} describing the gradual increase of absorption with decreasing wavelength ~\cite{khan2009charge}. It has been suggested that the 'ramp' could be related to vacancy clusters~\cite{maki2007effects,jones2009dislocations}, however, this can hardly explain the increase of the 'ramp' induced by e-beam irradiation, as electrons should not create significant multi-vacancy defects, at least for low fluences. Moreover, the 'ramp' fell down again after annealing (Fig.\ref{fig:Abs_Ann}), but the annealing temperature (1000~$^{\circ}$C) was below the dissociation point of vacancy clusters ($\sim$1500~$^{\circ}$C)~\cite{dobrinets2016hpht}. 
To conclude, the defect that increased the 'ramp' was induced proportional to the irradiation fluence, it can be largely annealed out at 1000~$^{\circ}$C, this suggests that single vacancies can contribute to the 'ramp'. 
But the exact cause of the 'ramp' and how it influences the NV formation still needs further investigation.
Furthermore, a center with ZPL at 489~nm (and a phonon side band centered at $\sim$460~nm) appeared after irradiation. The 489~nm center was attributed to a defect containing nitrogen atom bound to interstitial carbon atoms~\cite{field1992properties,collins1985absorption}, after annealing it vanished due to its low-temperature-stability~\cite{zaitsev2018defect}.

After annealing, vacancies were captured by P1 centers to form NV centers, resulting in a significant decrease in ND1 band and increase in NV band (ZPL at 637~nm for NV$^-$, phonon side band across to $\sim$450~nm), Fig.~\ref{fig:Abs_Ann}.
The ZPL for NV$^0$ (575~nm) is submerged in the broad side band of NV$^-$ for low fluences, but for the highest fluence (3E18~e/cm$^2$), due to the high concentration of NV$^0$, the ZPL for both NV$^-$ and NV$^0$ can be observed. 
For high fluences, isolated vacancies were not fully annealed out, which was particularly obvious for 3E18~e/cm$^2$ irradiation in Fig.~\ref{fig:Abs_Ann}b, that both ND1 and GR1 bands remained a considerable intensity. 
The reason can be either that the annealing temperature or duration was insufficient to convert all single vacancies, or more likely, P1 centers were much less available due to already high conversion, leading to a worse combination rate and a higher proportion of NV$^0$--which has been confirmed by the NV-PL result (Fig.~\ref{fig:Irr}c). 
The absorption and PL behavior of the series drove to the same conclusion, that an optimal fluence should be between 1E18 to 3E18~e/cm$^2$ in 1~MeV irradiation for diamonds with 2.2~ppm P1 centers.
This gives further support to our assumption that the appearance of GR1 centers after irradiation can indicate the NV$^0$ creation after annealing. 
Thus, the UV-Vis spectrum can help to decide if further irradiation is appropriate, for which we suggest it as a sufficient approach to determine the optimal irradiation fluence before annealing.
\subsubsection{NV creation for different initial P1 concentrations}
\label{chap:NDT_irradiated}
\begin{figure}[h!]
    \centering
    \begin{subfigure}[h]{0.97\linewidth}
        \centering
        \includegraphics[width=\linewidth]{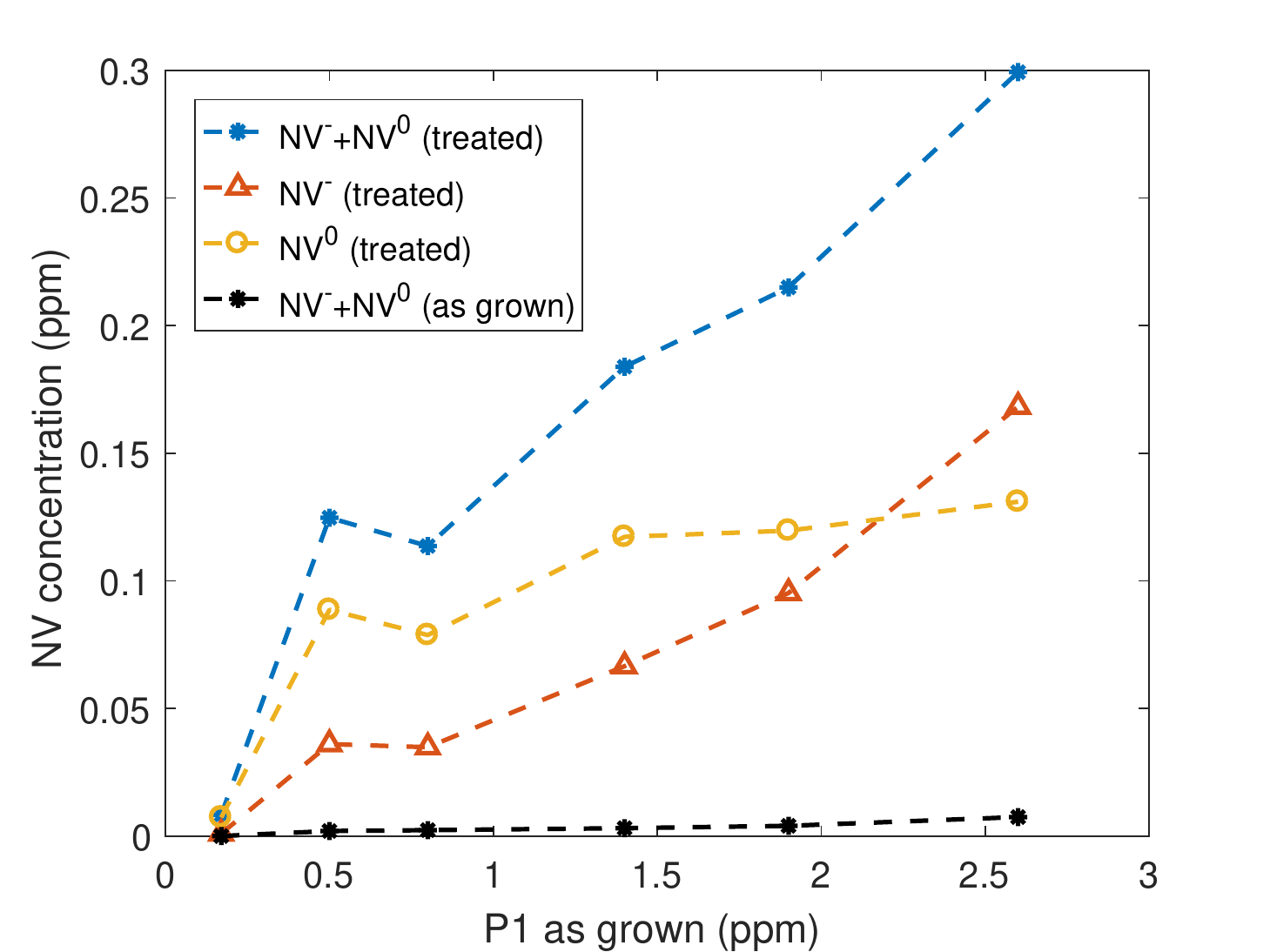}
        \put(-240,170){a)}
    \end{subfigure}
    \hfill
    \begin{subfigure}[h]{0.97\linewidth}
        \centering
        \includegraphics[width=\linewidth]{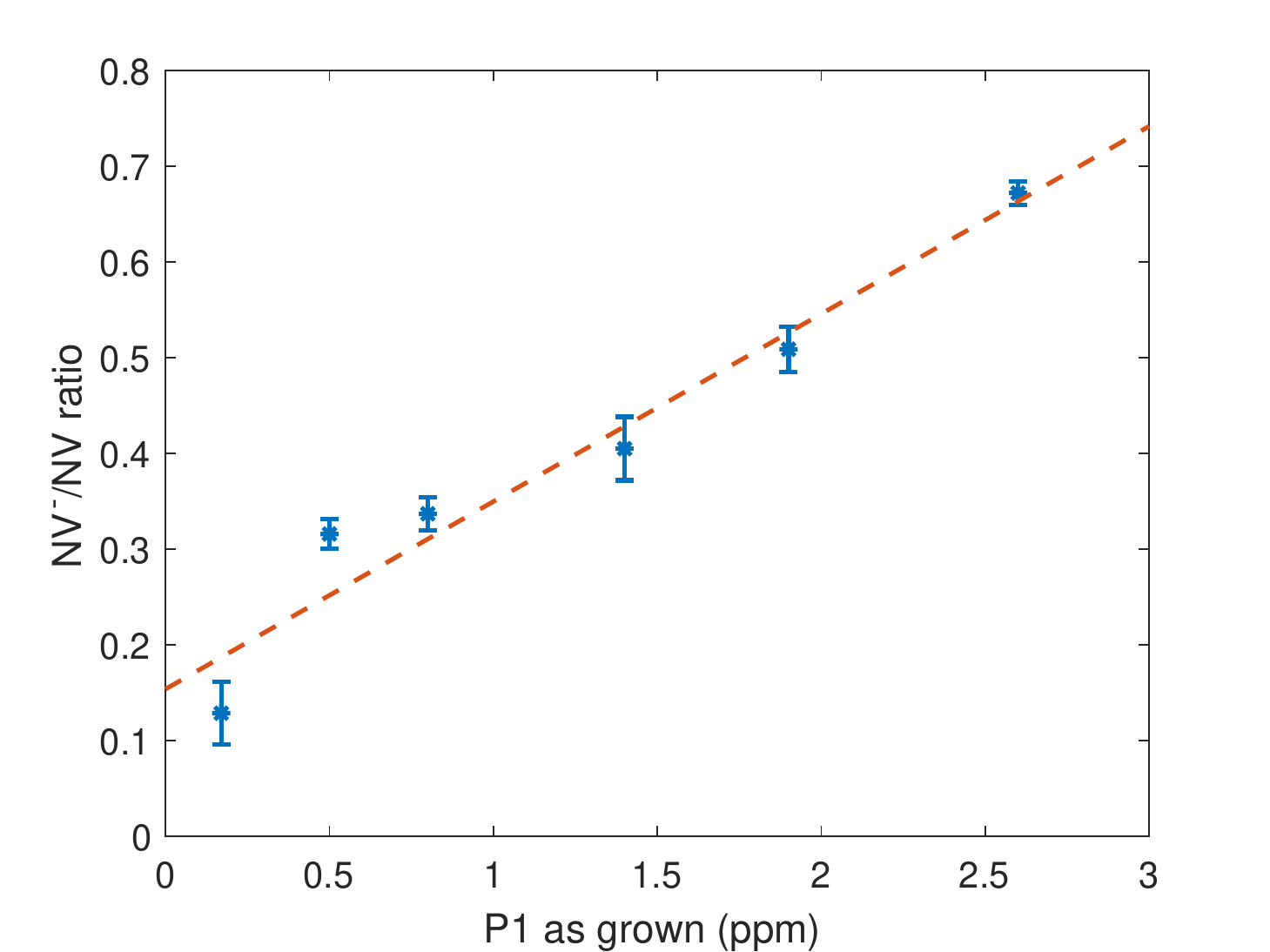}
        \put(-240,170){b)}
    \end{subfigure}
    \caption{The Nitrogen series~$\#$1 has been irradiated with 2~MeV, 2E17e/cm$^2$, which was optimized for $\sim$2.2~ppm P1 concentration. After irradiation and annealing, (a) the NV concentration and (b) the NV$^-$/NV ratio are both proportional to the as-grown P1 concentration. Errors are defined as in Fig~\ref{fig:P1NV_grown}.}
    \label{fig:NDT_treat}
\end{figure}

To study the NV creation from different as-grown P1 concentrations, as well as to confirm the optimal irradiation conditions, we irradiated the Nitrogen series~$\#$1 with P1 of 0.2-2.6~ppm from Sec.~\ref{chap:NV_asgrown}, with 2~MeV and used a fixed fluence of 2E17~e/cm$^2$. 
This fluence was chosen based on the discussion of the 2~MeV irradiation series in Sec.~\ref{chap:fluence}, for which a fluence of 1E17-2E17~e/cm$^2$ was the optimal regime for 2.2~ppm P1).
Fig.~\ref{fig:NDT_treat} illustrates the NV concentration and charge state ratio of the series after irradiation and annealing (also see Table.~\ref{table:NV_T2}). The NV creation was P1-limited under this irradiation condition, leading to a positive correlation between the NV and initial P1 concentration. 
The sample with 2.6 ppm P1 concentration shows the highest NV concentration and NV$^-$/NV ratio, as the fixed irradiation condition was optimized for a slightly lower P1 concentration of 2.2~ppm. 
This means the P1 concentration in this sample should be enough to charge the NV$^-$ center, comparing to the optimum in the irradiation series. 
In the end it contains 168~ppb NV$^-$ centers with 67$\%$ NV$^-$/NV ratio, $R^-_{con}\sim$~6.5$\%$. 
For the other samples with P1$<$2.6~ppm, more NV$^0$ presented due to over-irradiation, meaning the irradiation-induced vacancies converted too many P1 centers to NV centers, such that the remaining P1 concentration was no longer able to provide electron charges to the NV centers. 
Consequently, the NV$^-$/NV ratio (Fig.~\ref{fig:NDT_treat}b) showed a linear correlation to the initial P1 concentration, providing evidence that the optimal fluence can be positively correlated to the as-grown P1 concentration.
\subsubsection{Coherence time after irradiation and annealing}
\begin{figure}[h!]
    \centering
    \includegraphics[width=0.97\linewidth]{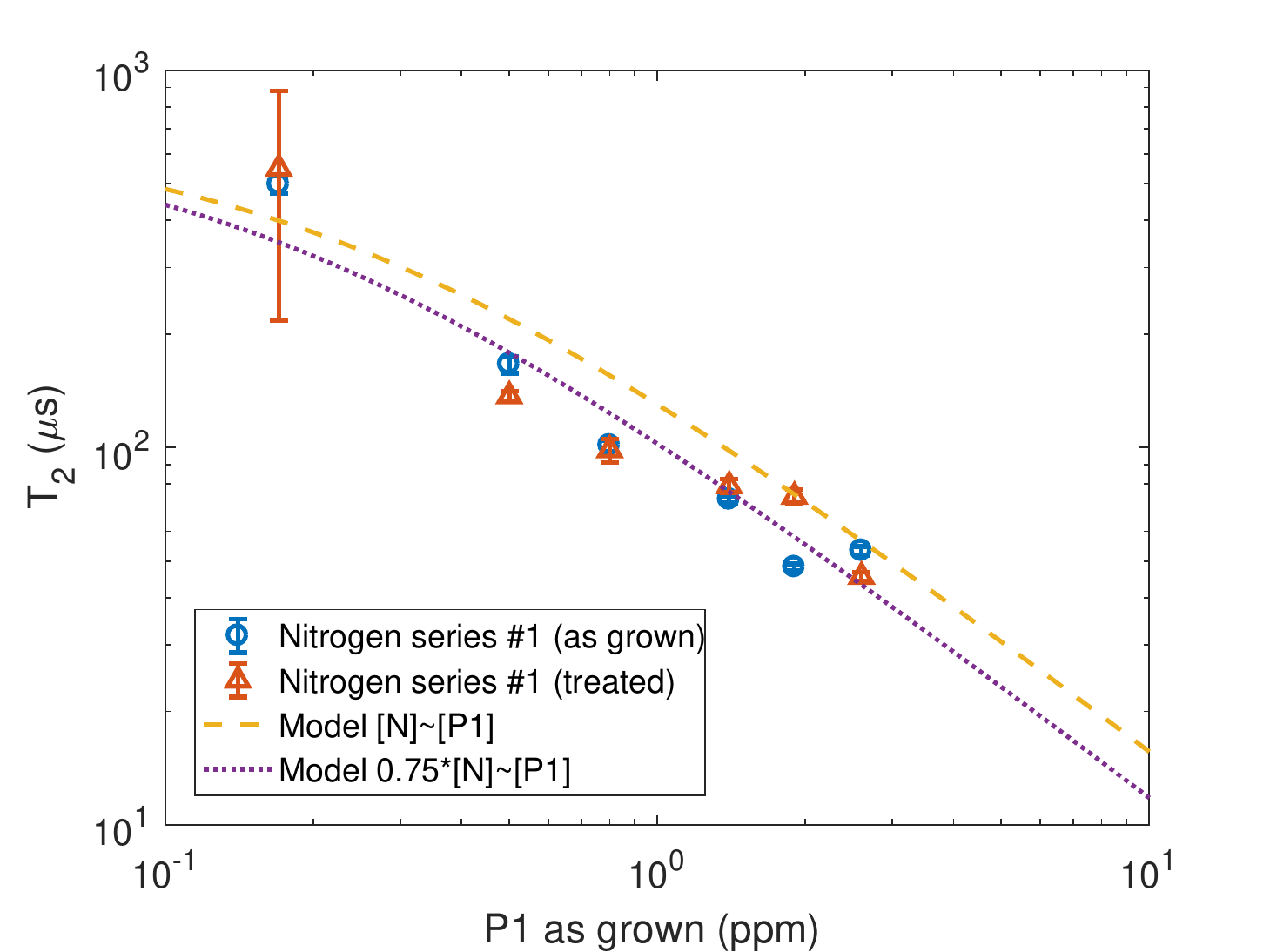}
    \caption{Coherence time before and after irradiation and annealing. after treatment, as the total nitrogen concentration in diamond is unchanged, the $T_2$ stays consistent (Details see also Table.~\ref{table:NV_T2}).}
    \label{fig:T2_treat}
\end{figure}

The irradiation and annealing have largely enhanced the NV concentration in our CVD series, leading to a stronger fluorescence signal, which can be the first approach to a better sensitivity according to Eq.~\ref{eq:sensitivity}. 
The coherence time, as another key factor, should be stable to prevent compromising the gained sensitivity. We assumed that the coherence time would stay consistent via the treatment, since the total nitrogen content in diamond would not be changed (Sec.~\ref{chap:T2grown}). 
As a confirmation, we measured $T_2$ of the Nitrogen series~$\#$1 (also in Fig.~\ref{fig:T2}a and Sec.~\ref{chap:NDT_irradiated}) again after treatment. 
Fig.~\ref{fig:T2_treat} shows that the $T_2$ stays approximately consistent, while the NV concentration increased significantly for the same samples (Fig.~\ref{fig:NDT_treat}a).

More details of NV$^-$ concentrations and $T_2$ times of the series are listed in Table.~\ref{table:NV_T2}. 
After irradiation and annealing, the series obtains an NV$^-$ concentration of 1-168~ppb with a $T_2$ time of 46-549~$\mu$s.
The longest $T_2$ time we achieved, 549~$\mu$s (NDT-26), is comparable to previously reported $T_2$=600~$\mu$s for NV ensembles~\cite{stanwix2010coherence} in high purity diamonds (with a natural abundance of 1.1$\%$ $^{13}$C), but our NV concentration is higher (1~ppb NV$^-$ centers in this work vs. $\sim$0.18~ppb NV centers in Stanwix~{\em et al.}). Our longest $T_2$ time is also close to reported $T_2$ times for single NV of $\sim$600-650~$\mu$s~\cite{maze2008nanoscale, mizuochi2009coherence}.

\begin{table*}[h!]
\footnotesize
\centering
 \begin{tabular}{M{4em} M{4em} M{5em} M{5em} M{5em} M{5em} M{7em} M{6em} M{6em}} 
  \toprule
  Series & Sample & N/C ratio (ppm) & As-grown P1 (ppm) & As-grown NV$^-$ (ppb) & Treated NV$^-$ (ppb) & Treated NV$^-$/NV (\%) & As-grown $T_2$ ($\mu$s) & Treated $T_2$ ($\mu$s)\\
  \hline
  \multirow{7}{4em}{Nitrogen series $\#$1}& NDT-26 & 150 & 0.2 & 0.03 & 1 & 12.8 $\pm$ 1.7 &  497.7 $\pm$ 26.2 & 549 $\pm$ 332\\
  & NDT-14 & 500 & - & 0.2 & 10 & 38.4 $\pm$ 3.3 & 288.9 $\pm$ 31.3 & 329.3 $\pm$ 101.8\\ 
  & NDT-07 & 1500 & 0.5 & 1.5 & 36 & 31.6 $\pm$ 1.6 & 166.1 $\pm$ 8.7 & 136.9 $\pm$ 3.9\\ 
  & NDT-34 & 2500 & 0.8 & 1.8 & 35 & 33.7 $\pm$ 3.0 & 101.3 $\pm$ 3.3 & 98.5 $\pm$ 7.1\\ 
  & NDT-01 & 4500 & 1.4 & 2.4 & 67 & 40.5 $\pm$ 3.3 & 72.8 $\pm$ 1.5 & 79.1 $\pm$ 3.2\\ 
  & NDT-02 & 7429 & 1.9 & 3.3 & 95 & 50.9 $\pm$ 2.3 & 48.2 $\pm$ 1.0 & 74.2 $\pm$ 3.4\\ 
  & NDT-12 & 8500 & 2.6 & 6.4 & 168 & 67.2 $\pm$ 1.2 & 53.3 $\pm$ 1.4 & 45.5 $\pm$ 1.2\\ 
  \midrule
  \multirow{7}{4em}{Nitrogen series $\#$2} & Cas-40 & 9722& 3.2 & 15.3 & \\ 
  & Cas-48 & 42777 & 5.2 & 21.3 & \\ 
  & Cas-68 & 77142 & 7.8 & 16.0 & \\ 
  & Cas-44 & 87499 & 9.5 & 25.9 & \\ 
  & Cas-51 & 173571 & 11.2 & 33.9 & \\ 
  & Cas-49 & 347143 & 13.0 & 27.1 & \\ 
  & Cas-50 & 694286 & 19.3 & 28.5 & \\
  \cmidrule[1pt](lr){1-5}
 \end{tabular}
\caption{NV creations and coherence times in nitrogen-varying CVD series. The 'treated' data were acquired after irradiation with 2~MeV, 1E17~e/cm$^2$, then annealing at 1000~$^{\circ}$C for 2h. Errors are defined as in Fig.~\ref{fig:P1NV_grown} and Fig.~\ref{fig:T2}.}
\label{table:NV_T2}
\end{table*}

The material-limited sensitivity is difficult to define, previously Edmonds~{\em et al.}~\cite{edmonds2021characterisation} has discussed it through the product of P1 concentration and coherence time. 
Considering that the coherence time is restricted by the P1 concentration, the product of them has an upper boundary which limits the sensitivity in theory.
In practice, according to Eq.~\ref{eq:sensitivity}, the product of NV fluorescence signal $I$ and the coherence time $\tau$ plays a more direct role.
Since P1 centers can be only partially converted to NV$^-$ centers, the theoretical limit by the material is unreachable, however, enhancing the P1 to NV$^-$ conversion rate can push the sensitivity closer to the limit. For our Nitrogen series $\#$1, the product of NV$^-$ and T$_2$ increased from 20-70 times though optimizing the e-beam irradiation, leading to an increase in the $\sqrt{I\cdot\tau}$ factor of around 5-8 times.

On the other hand, the ODMR contrast $C$ is equally important for the sensitivity, calling for a high NV$^-$/NV ratio.
In the Nitrogen series $\#$1, as discussed in Sec.~\ref{chap:NDT_irradiated}, the sample NDT-12 with 2.6~ppm P1 shows the highest NV$^-$/NV ratio after irradiation and annealing steps. All other samples with lower nitrogen concentrations--although they have longer T$_2$--contain high proportions of NV$^0$ centers.
By adapting irradiation fluences for corresponding P1 concentrations, high NV$^-$ concentrations and NV$^-$/NV ratios can be both satisfied. That, coupled with the optimization of the growth for long $T_2$ times, provides possibilities to further improve the sensitivity, by modulating the combination of NV$^-$ concentrations and $T_2$ times.

\section{Conclusion}
We investigated nitrogen-doped CVD-grown diamond by varying the N/C ratio over a range from 150 to 10$^6$~ppm.
We found that neutral single substitutional nitrogen (P1 centers) are created with a dependence of $\sim0.09\sqrt{ \text{N/C}}$ and created P1 densities in the CVD growth from 0.2 to 20~ppm. 
NV centers were also created during the growth with a fixed ratio of NV$^-$/P1 density of 0.25$\%$. The NV densities after growth were 0.03-33.9~ppb. 
Coherence times T$_2$ of the NV center after growth showed the expected inverse relationship with P1 centers, and they spanned 48-497~$\mu$s for P1 concentrations at the range of 0.2-2.6~ppm.
For that range, we showed the coherence time as a function of the P1 concentration. 
Comparing this to a previous study where absolute nitrogen content was studied by SIMS, we found agreement if we assume that ~75$\%$ of the nitrogen was in the form of P1 centers. 
The created range of NV densities and coherence times showed that as-grown nitrogen-doped CVD diamond can be used as a sufficient and reproducible sensing material.

We investigated and optimized the fluence of subsequent high-energy electron irradiation. 
We found that increasing irradiation fluence increases NV concentrations, however, there is a certain optimal point, above which more fluence creates mainly additional NV$^0$ and hardly any more NV$^-$. 
This optimum for 2.2~ppm initial P1 was at 1E17-2E17~e/cm$^2$ for 2~MeV and at 1E18-3E18~e/cm$^2$ for 1MeV irradiation. 
With this optimum, we achieved a P1 to NV$^-$ conversion rate $\sim$7-8$\%$, with an NV$^-$/NV ratio $\sim$66-86$\%$.
Since P1 centers are the main electron donors to charge the NV centers, we interpret the creation of NV$^0$ above the optimum as the conversion of too many P1 centers to NV centers, such that not enough P1 centers are left to provide electrons to charge the NV center.
Based on this model, we assume that the optimal fluence scales with the P1 density for similarly grown CVD diamonds.
We also found that for CVD diamonds, the conversion rate from as-grown P1 to NV centers should be smaller than 10$\%$ to fulfill an NV$^-$/NV ratio above 80$\%$.
In the future, this might be overcome by co-doping with another electron donor, for example phosphorous as a potential candidate.

Furthermore, we studied the absorption spectral behavior at the UV-Visible range to visually describe defect transformations during irradiation and annealing. 
We observed the generation of V$^-$ centers (ND1 band) after irradiation, scaling with the fluence. 
Neutral V$^0$ centers (GR1 band) were also formed for high fluences.
After annealing, V$^-$ centers were largely converted, forming NV centers proportional to the irradiation fluence.
We found that the appearance of the GR1 band after irradiation can be an indicator of the NV$^0$ formation after annealing. 
This supports further to our model, that for over-irradiated samples, the remaining P1 centers are not enough to charge vacancies (or NV centers), therefore neutral vacancies and later neutral NV centers tend to form.
From that, we provided a novel approach from the UV-Vis absorption spectrum to determine the optimal irradiation fluence, before annealing the sample.

Being treated with our irradiation and annealing protocol, the coherence time T$_2$ stay consistent. 
This means the irradiation/annealing provides significant advantages for sensing and sensitivity by increasing the NV density and thus the signal strength without compromising the T$_2$. 
After treatment, we achieved a combination between 549~$\mu$s T$_2$ with 1~ppb NV$^-$ and 45.5~$\mu$s with 168~ppb.
The longest T$_2$=549~$\mu$s is comparable to T$_2$ reported previously for single NV~\cite{maze2008nanoscale, mizuochi2009coherence}, as well as for NV ensembles~\cite{stanwix2010coherence}, but with higher NV$^-$ concentration in our sample. 
Combining the enhanced NV$^-$ concentration with long T$_2$ is an interesting pathway to improve sensitivities in sensing.
\section{Acknowledgement}
We thank Xavier Vidal, Arne Götze, Oliver Ambacher, Brant Gibson, Philipp Reineck and Andrew Greentree for valuable discussions.
We thank Michael Ardner for cutting and polishing the diamond plates, Dorothee Luick for the technical support of UV-Vis measurements, Shangjing Liu for the technical support of PL measurements, and Sioe See Volaric for technical EPR assistance.
The part of electron irradiation was carried out within the framework of QST Internal Research Initiative.  
T.L and J.J acknowledge the funding by the German federal ministry for education and research Bundesministerium für Bildung und Forschung (BMBF) under Grant No. 13XP5063. 
B.C.J. acknowledges support from the ARC Centre of Excellence for Quantum Computation and Communication Technology (CE170100012).
D.W. acknowledges the support through an Australian Government Research Training Program Scholarship.
M.C. acknowledges funding from the Asian Office of Aerospace Research and Development (FA2386-18-1-4056).


\bibliographystyle{elsarticle-num}
\bibliography{references}
\end{document}